\documentclass[10pt,journal,compsoc]{IEEEtran}
\ifCLASSOPTIONcompsoc
  \usepackage[nocompress]{cite}
\else
  \usepackage{cite}
\fi
\ifCLASSINFOpdf
\else
\fi

\usepackage[hyphens]{url}
\usepackage{filecontents}
\usepackage{graphicx}
\usepackage[font=small,labelfont=bf]{caption}
\usepackage[utf8]{inputenc}
\usepackage{soul}
\usepackage{balance}
\usepackage{glossaries}
\usepackage{xcolor}
\usepackage{comment}
\usepackage{subcaption}
\usepackage{cleveref}

\newcommand{\mI}{\mathcal{I}}
\newcommand{\mJ}{\mathcal{J}}
\usepackage[inline]{enumitem}
\setacronymstyle{long-short}
\newacronym{quic}{QUIC}{Quick UDP Internet Connections}
\newacronym{3gpp}{3GPP}{3rd Generation Partnership Project}
\newacronym{adc}{ADC}{Analog to Digital Converter}
\newacronym{5g}{5G}{5th Generation}
\newacronym{aimd}{AIMD}{Additive Increase Multiplicative Decrease}
\newacronym{am}{AM}{Acknowledged Mode}
\newacronym{amc}{AMC}{Adaptive Modulation and Coding}
\newacronym{aqm}{AQM}{Active Queue Management}
\newacronym{awgn}{AGWN}{Additive White Gaussian Noise}
\newacronym{balia}{BALIA}{Balanced Link Adaptation}
\newacronym{bdp}{BDP}{Bandwidth-Delay Product}
\newacronym{bf}{BF}{Beamforming}
\newacronym{cc}{CC}{Congestion Control}
\newacronym{cdf}{CDF}{Cumulative Distribution Function}
\newacronym{cn}{CN}{Core Network}
\newacronym{cqi}{CQI}{Channel Quality Information}
\newacronym{cp}{CP}{Control Plane}
\newacronym{csirs}{CSI-RS}{Channel State Information - Reference Signal}
\newacronym{dc}{DC}{Dual Connectivity}
\newacronym{dce}{DCE}{Direct Code Execution}
\newacronym{dci}{DCI}{Downlink Control Information}
\newacronym{dl}{DL}{Downlink}
\newacronym{dmr}{DMR}{Deadline Miss Ratio}
\newacronym{dmrs}{DMRS}{DeModulation Reference Signal}
\newacronym{e2e}{E2E}{End-to-End}
\newacronym{ecn}{ECN}{Explicit Congestion Notification}
\newacronym{edf}{EDF}{Earliest Deadline First}
\newacronym{enb}{eNB}{evolved Node Base}
\newacronym{epc}{EPC}{Evolved Packet Core}
\newacronym{es}{ES}{Edge Server}
\newacronym{fdma}{FDMA}{Frequency Division Multiple Access}
\newacronym{fdd}{FDD}{Frequency Division Duplexing}
\newacronym[firstplural=Radio Access Technologies (RATs)]{rat}{RAT}{Radio Access Technology}
\newacronym{fs}{FS}{Fast Switching}
\newacronym{ftp}{FTP}{File Transfer Protocol}
\newacronym{gnb}{gNB}{Next Generation Node Base}
\newacronym{harq}{HARQ}{Hybrid Automatic Repeat reQuest}
\newacronym{hetnet}{HetNet}{Heterogeneous Network}
\newacronym{hh}{HH}{Hard Handover}
\newacronym{hol}{HOL}{Head-of-Line}
\newacronym{ia}{IA}{Initial Access}
\newacronym{imt}{IMT}{International Mobile Telecommunication}
\newacronym{iot}{IoT}{Internet of Things}
\newacronym{los}{LOS}{Line of Sight}
\newacronym{lte}{LTE}{Long Term Evolution}
\newacronym{m2m}{M2M}{Machine to Machine}
\newacronym{mac}{MAC}{Medium Access Control}
\newacronym{mc}{MC}{Multi-Connectivity}
\newacronym{mcs}{MCS}{Modulation and Coding Scheme}
\newacronym{mec}{MEC}{Mobile Edge Cloud}
\newacronym{mi}{MI}{Mutual Information}
\newacronym{mimo}{MIMO}{Multiple Input, Multiple Output}
\newacronym{mmwave}{mmWave}{millimeter wave}
\newacronym{mr}{MR}{Maximum Rate}
\newacronym{mss}{MSS}{Maximum Segment Size}
\newacronym{mtd}{MTD}{Machine-Type Device}
\newacronym{mtu}{MTU}{Maximum Transmission Unit}
\newacronym{nsf}{NSF}{National Science Foundation}
\newacronym{nfv}{NFV}{Network Function Virtualization}
\newacronym{nlos}{NLOS}{Non Line of Sight}
\newacronym{nr}{NR}{New Radio}
\newacronym{ofdm}{OFDM}{Orthogonal Frequency Division Multiplexing}
\newacronym{pdcch}{PDCCH}{Physical Downlonk Control Channel}
\newacronym{pdcp}{PDCP}{Packet Data Convergence Protocol}
\newacronym{pdsch}{PDSCH}{Physical Downlink Shared Channel}
\newacronym{pdu}{PDU}{Packet Data Unit}
\newacronym{pf}{PF}{Proportional Fair}
\newacronym{pgw}{PGW}{Packet Gateway}
\newacronym{phy}{PHY}{Physical}
\newacronym{pbch}{PBCH}{Physical Broadcast Channel}
\newacronym[plural=\gls{mme}s,firstplural=Mobility Management Entities (MMEs)]{mme}{MME}{Mobility Management Entity}
\newacronym{prb}{PRB}{Physical Resource Block}
\newacronym{pss}{PSS}{Primary Synchronization Signal}
\newacronym{pucch}{PUCCH}{Physical Uplink Control Channel}
\newacronym{pusch}{PUSCH}{Physical Uplink Shared Channel}
\newacronym{rach}{RACH}{Random Access Channel}
\newacronym{ran}{RAN}{Radio Access Network}
\newacronym{red}{RED}{Random Early Detection}
\newacronym{rf}{RF}{Radio Frequency}
\newacronym{rlc}{RLC}{Radio Link Control}
\newacronym{rlf}{RLF}{Radio Link Failure}
\newacronym{rrc}{RRC}{Radio Resource Control}
\newacronym{rrm}{RRM}{Radio Resource Management}
\newacronym{rr}{RR}{Round Robin}
\newacronym{rs}{RS}{Remote Server}
\newacronym{rsrp}{RSRP}{Reference Signal Received Power}
\newacronym{rss}{RSS}{Received Signal Strength}
\newacronym{rtt}{RTT}{Round Trip Time}
\newacronym{rw}{RW}{Receive Window}
\newacronym{rx}{RX}{Receiver}
\newacronym{sa}{SA}{standalone}
\newacronym{sack}{SACK}{Selective Acknowledgment}
\newacronym{sap}{SAP}{Service Access Point}
\newacronym{sch}{SCH}{Secondary Cell Handover}
\newacronym{scoot}{SCOOT}{Split Cycle Offset Optimization Technique}
\newacronym{sdma}{SDMA}{Spatial Division Multiple Access}
\newacronym{sinr}{SINR}{Signal-to-Interference-plus-Noise-Ratio}
\newacronym{sm}{SM}{Saturation Mode}
\newacronym{snr}{SNR}{Signal-to-Noise-Ratio}
\newacronym{son}{SON}{Self-Organizing Network}
\newacronym{ss}{SS}{Synchronization Signal}
\newacronym{srs}{SRS}{Sounding Reference Signal}
\newacronym{sss}{SSS}{Secondary Synchronization Signal}
\newacronym{tb}{TB}{Transport Block}
\newacronym{tcp}{TCP}{Transmission Control Protocol}
\newacronym{tdd}{TDD}{Time Division Duplexing}
\newacronym{tdma}{TDMA}{Time Division Multiple Access}
\newacronym{tfl}{TfL}{Transport for London}
\newacronym{thz}{THz}{Terahertz}
\newacronym{tm}{TM}{Transparent Mode}
\newacronym{trp}{TRP}{Transmitter Receiver Pair}
\newacronym{tti}{TTI}{Transmission Time Interval}
\newacronym{ttt}{TTT}{Time-to-Trigger}
\newacronym{tx}{TX}{Transmitter}
\newacronym{ue}{UE}{User Equipment}
\newacronym{ul}{UL}{Uplink}
\newacronym{uml}{UML}{Unified Modeling Language}
\newacronym{um}{UM}{Unacknowledged Mode}
\newacronym{utc}{UTC}{Urban Traffic Control}
\newacronym{vm}{VM}{Virtual Machine}
\newacronym{rsrq}{RSRQ}{Reference Signal Received Quality}
\newacronym{rssi}{RSSI}{Received Signal Strength Indicator}
\newacronym{crs}{CRS}{Cell Reference Signal}
\newacronym{comp}{CoMP}{Coordinated Multi-Point}
\newacronym{cran}{C-RAN}{Cloud \acrlong{ran}}
\newacronym{ca}{CA}{Carrier Aggregation}
\newacronym{cco}{CC}{Carrier Component}
\newacronym{nsa}{NSA}{Non Stand Alone}
\newacronym{embb}{eMBB}{Enhanced Mobility Broadband}
\newacronym{bsr}{BSR}{Buffer Status Report}
\newacronym{srb}{SRB}{Service Radio Bearer}
\newacronym{scm}{SCM}{Spatial Channel Model}
\newacronym{sctp}{SCTP}{Stream Control Transmission Protocol}
\newacronym{mptcp}{MPTCP}{Multi-path TCP}
\newacronym{ietf}{IETF}{Internet Engineering Task Force}
\newacronym{os}{OS}{Operating System}
\newacronym{tls}{TLS}{Transport Layer Security}
\newacronym{rfc}{RFC}{Request for Comments}
\newacronym{http}{HTTP}{HyperText Transfer Protocol}
\newacronym{nat}{NAT}{Network Address Translation}
\newacronym{api}{API}{Application Programming Interface}
\newacronym{rto}{RTO}{Retransmission Timeout}
\newacronym{psc}{PSC}{Public Safety Communication}
\newacronym{rpgm}{RPGM}{Reference Point Group Mobility}
\newacronym{ic}{IC}{Incident Command}
\newacronym{rsu}{RSU}{Road Side Unit}
\newacronym{usa}{U.S.}{United States}
\newacronym{vr}{VR}{Virtual Reality}
\newacronym{iab}{IAB}{Integrated Access and Backhaul}
\newacronym{wlan}{WLAN}{Wireless Local Area Network}
\newacronym{cots}{COTS}{Commercial Off-the-Shelf}
\newacronym{fpga}{FPGA}{Field Programmable Gate Array}
\newacronym{rcn}{RCN}{Research Coordination Network}
\newacronym{isp}{ISP}{Internet Service Provider}
\newacronym{jfi}{JFI}{Jain's Fairness Index}
\newacronym{udp}{UDP}{User Datagram Protocol}
\newacronym{qos}{QoS}{Quality of Service}
\newacronym{ncp}{NCP}{Network Control Problem}
\newacronym{npsd}{NPSD}{Noise Power Spectral Density}
\newacronym{srslte}{srsLTE}{srsLTE}
\newacronym{ici}{ICI}{Inter-cell Interference}
\newacronym{bs}{BS}{Base Station}
\newacronym{scofdma}{SC-OFDMA}{Single Carrier Orthogonal Frequency-division Multiple Access}
\newacronym{cpofdm}{CP-OFDM}{Cyclic Prefix Orthogonal Frequency-division Multiplexing}
\newacronym{dftsofdm}{DFT-s-OFDM}{Discrete Fourier Transform Spread Orthogonal Frequency Division Multiplexing}
\newacronym{sdr}{SDR}{Software-Defined Radio}
\newacronym{usrp}{USRP}{Universal Software Radio Peripheral}
\newacronym{uav}{UAV}{Unmanned Aerial Vehicle}
\newacronym{uas}{UAS}{Unmanned Aerial System}
\newacronym{poi}{POI}{Point of Interest}

\begin{document}

\newcommand{\ft}[1]{$#1\:\mathrm{ft}$}
\newcommand{\meter}[1]{$#1\:\mathrm{m}$}
\newcommand{\metersec}[1]{$#1\:\mathrm{m/s}$}
\newcommand{\ghz}[1]{$#1\:\mathrm{GHz}$}
\newcommand{\mhz}[1]{$#1\:\mathrm{MHz}$}
\newcommand{\db}[1]{$#1\:\mathrm{dB}$}
\newcommand{\dbm}[1]{$#1\:\mathrm{dBm}$}
\newcommand{\eNB}[1]{eNB\:$#1$}
\newcommand{\ue}[1]{UE\:$#1$}
\newcommand{\eg}{e.g., }
\newcommand{\ie}{i..e, }
\newcommand{\fix}{ \textbf{ \color{red} [XX] } }
\newcommand{\reworded}[1]{{\color{black}#1}}
\newcommand{\new}[1]{{\color{black}#1}}
\newcommand{\todo}[1]{\textbf{\newline \color{red}[ #1 ]}\newline}
\newcommand{\mbps}[1]{\:\mathrm{Mbps}#1}
\newcommand{\redo}[1]{\textbf{\newline \color{blue}[ #1 ]}\newline}
\newcommand{\red}[1]{\textcolor{red}{#1}}
\newcommand{\bb}[1]{\textcolor{blue}{#1}}

%
\title{\textit{Streaming from the Air}: Enabling Drone-sourced Video Streaming Applications on 5G Open-RAN Architectures}
%
%
%
%

\author{Lorenzo Bertizzolo,
        Tuyen X. Tran,~\IEEEmembership{Member,~IEEE,}
        John Buczek,
        Bharath Balasubramanian,~\IEEEmembership{Member,~IEEE,}
        Rittwik Jana,~\IEEEmembership{Member,~IEEE,}
        Yu Zhou, 
        and~Tommaso Melodia,~\IEEEmembership{Fellow,~IEEE}

%
\thanks{This work has been submitted to the IEEE for possible publication.
Copyright may be transferred without notice, after which this version may no longer be accessible.}
}

\IEEEtitleabstractindextext{%
\begin{abstract}
Enabling high data-rate uplink cellular connectivity for drones is a challenging problem, since
a flying drone has a higher likelihood of having line-of-sight propagation to base stations that terrestrial UEs normally do not have line-of-sight to.
This may result in uplink inter-cell interference and uplink performance degradation for the neighboring ground UEs when drones transmit at high data-rates (\eg video streaming).
We address this problem from a cellular operator's standpoint to support drone-sourced video streaming of a point of interest. We propose a low-complexity, closed-loop control system for Open-RAN architectures that jointly optimizes the drone's location in space and its transmission directionality to support video streaming and minimize its uplink interference impact on the network. 
We prototype and experimentally evaluate the proposed control system on a dedicated outdoor multi-cell RAN testbed, which is the first measurement campaign of its kind. Furthermore, we perform a large-scale simulation assessment of the proposed control system using the actual cell deployment topologies and cell load profiles of a major US cellular carrier. 
The proposed Open-RAN control scheme achieves an average $19\%$ network capacity gain over traditional BS-constrained control solutions and satisfies the application data-rate requirements of the drone (\eg to stream an HD video).
\end{abstract}

}

\maketitle

\IEEEdisplaynontitleabstractindextext

%
\IEEEpeerreviewmaketitle

\IEEEraisesectionheading{\section{Introduction}\label{sec:introduction}}
Drones \cite{buczek2021wireless} (or Unmanned Aerial Vehicles, ``UAVs'')-sourced video streaming enables a wide range of new services and applications such as aerial surveillance, infrastructure monitoring, environmental sensing, transportation and delivery of goods, and live broadcast coverage~\cite{naqvi2018drone, boubrima2020robust, shell, amazon, cnn}. 
Employing drones for these applications can significantly lower the chance for human operators' injuries and reduce the overall operational cost when compared to larger form-factor solutions such as helicopters. To support beyond visual-line-of-sight control, drones will be equipped with on-board cellular user-equipment (UEs). Thanks to the cellular network coverage and support for broadband data-rates, drones will not require a flight operator nearby, but instead will be remotely operated over the Internet at unprecedented distances. At the same time, they will be able to connect to remote broadcast servers and upload live video streams, such as live footage captured via on-board cameras, over the air-to-ground (A2G) cellular links.
\begin{figure*}[t!]
  \centering
  \subcaptionbox{
  UAV as a Ground UE.
\label{fig:motivationNLos}
  }
  {
    \includegraphics[width=.53\columnwidth]
    {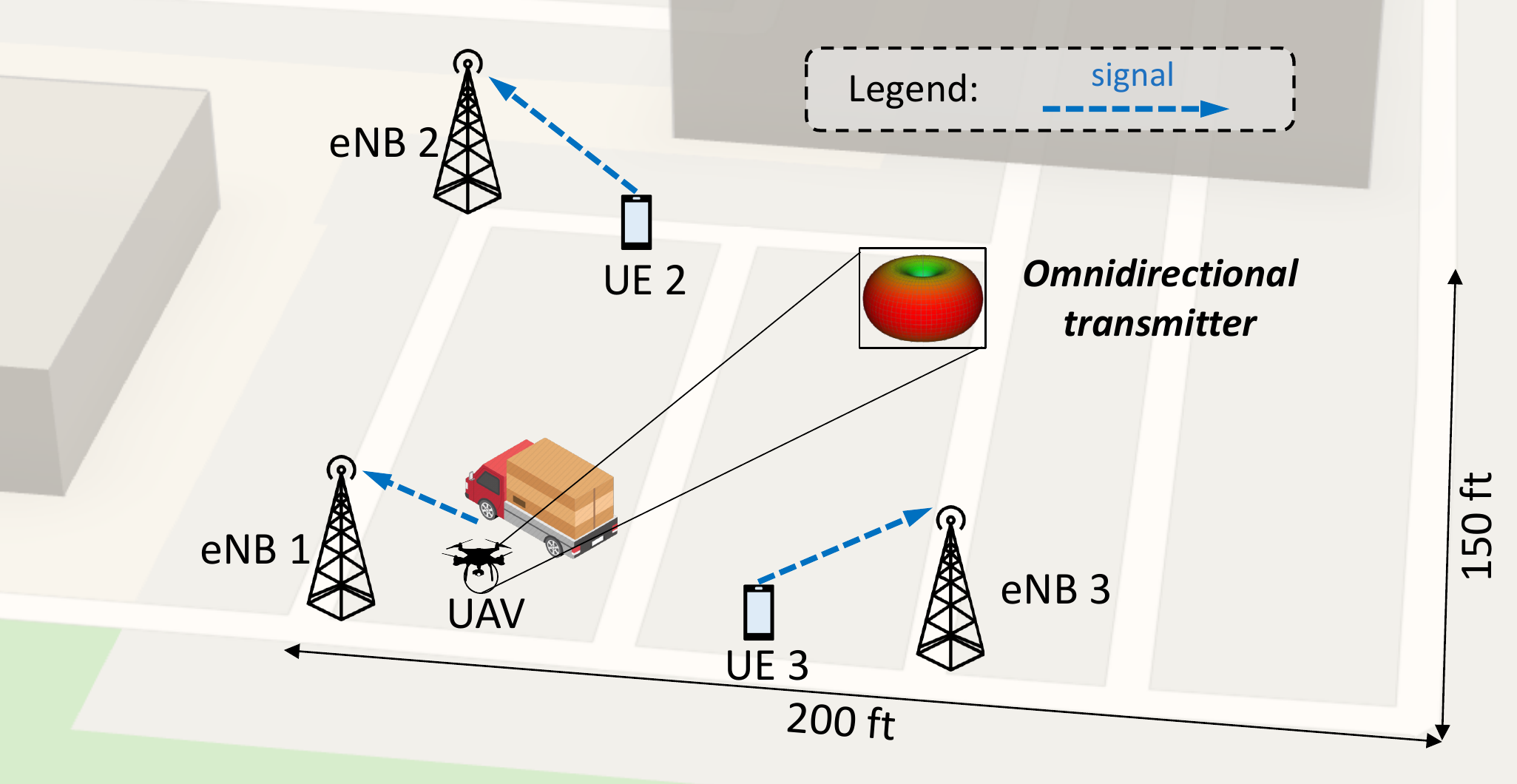}
  }
  \subcaptionbox{
    UAV  with omnidirectional TX.
  \label{fig:motivationLos}
  }
  {
    \includegraphics[width=.53\columnwidth]
    {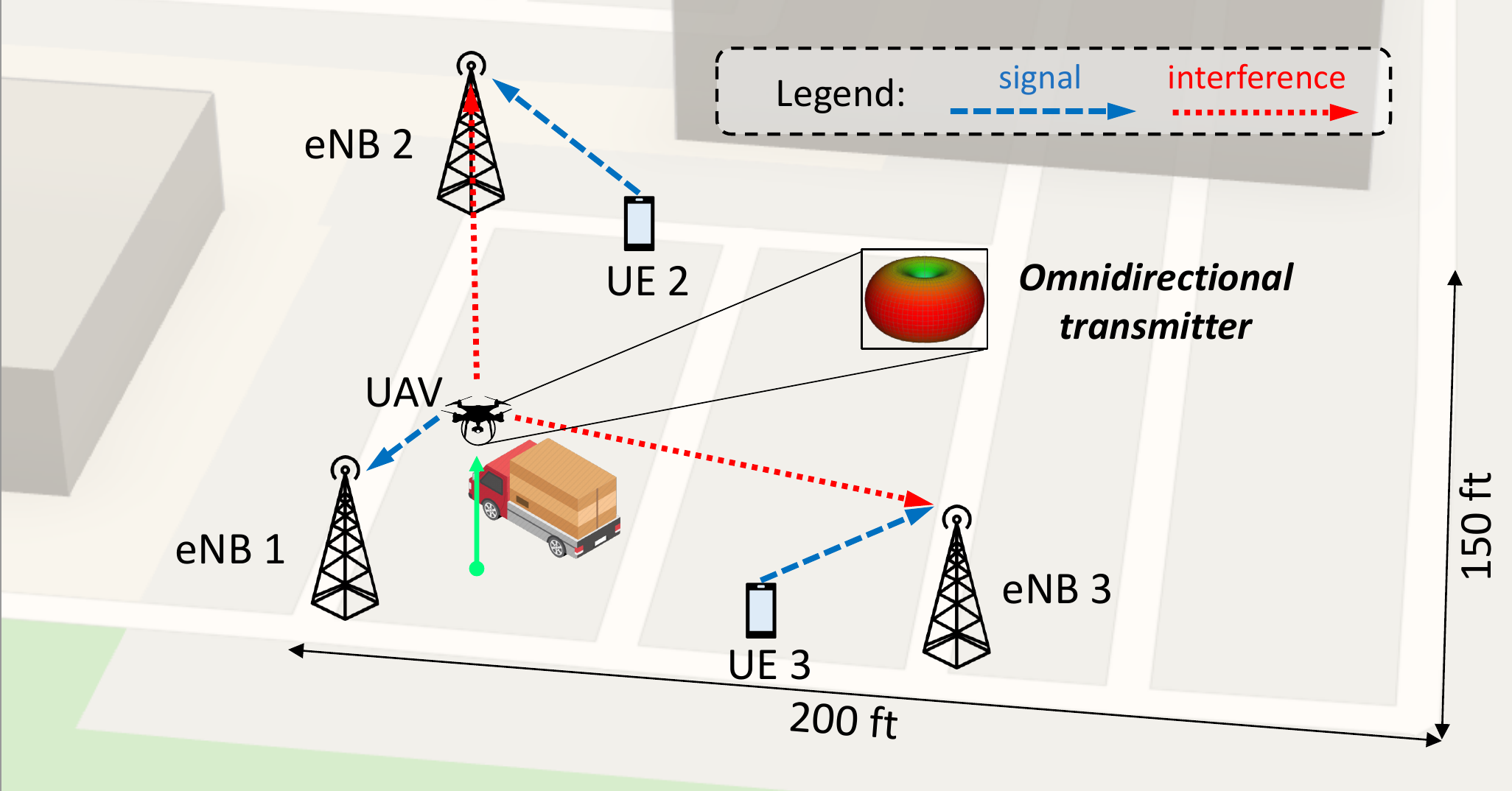}
  }
  \subcaptionbox{
    UAV with directional TX.
  \label{fig:motivationLosDir}
  }
  {
    \includegraphics[width=.53\columnwidth]
    {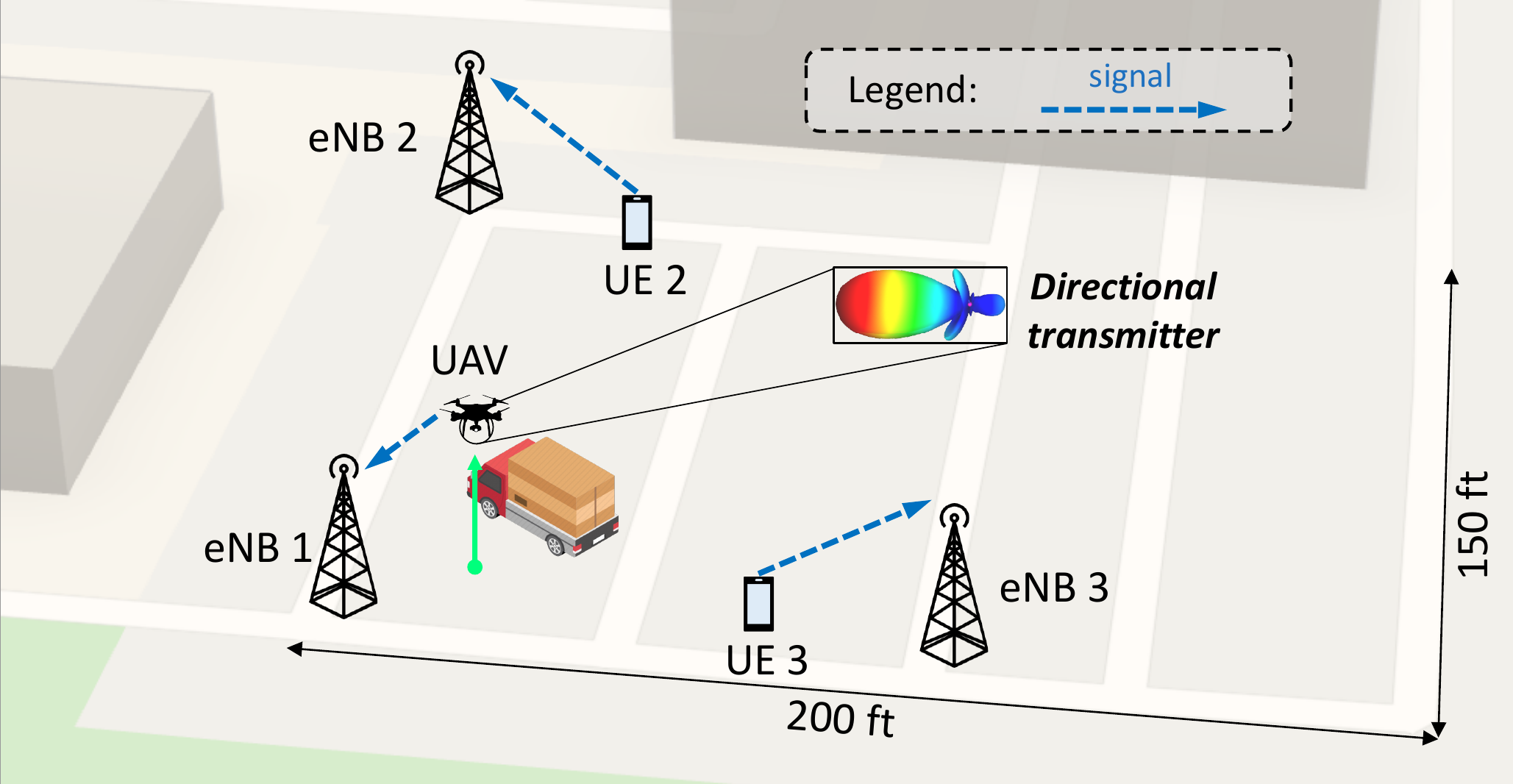}
  }
  \subcaptionbox{
    Performance.
    \label{fig:motivationPerformance}
  }
  {
    \includegraphics[width=.3\columnwidth]
    {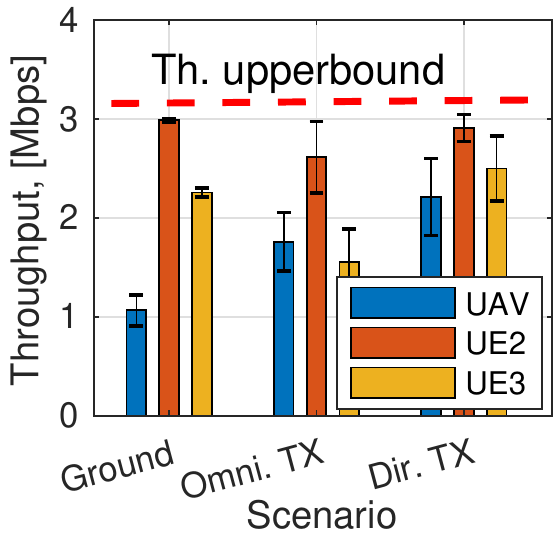}
  }
  \caption{
  Motivational network deployment: 
  $3$-eNBs serving $2$ UEs and one UAV-based UE employing different antenna patterns.} \vspace{0mm}
  \label{fig:motivation_topology}
\end{figure*}
Unlike traditional terrestrial UEs that are typically situated at pedestrian heights, UAV-based UEs will operate at high altitudes and therefore in LoS condition with multiple (serving and neighboring) cellular base stations (BSs). 
While this condition is favorable for the drone's uplink traffic \cite{BertizzoloHotmobile20}, the drone's transmitted uplink signal will also propagate to multiple neighboring BSs, resulting in severe uplink inter-cell interference and uplink performance degradation for the ground UEs served by these BSs~\cite{amorim2018measured}. 
This problem is exacerbated when the drones transmit at high data-rates, for example, when streaming HD videos. 
As drone-sourced video streaming is a major application of interest, industries are posing significant pressure on network operators to support high bandwidth drone uplink communications in future 5G-and-beyond networks \cite{3gpprel15, vodafone}. 

The need for drone cellular support in FR1 was first introduced by the 3GPP in release 15 and carried on successive releases for operations in the 5G NR low- and mid-band (\ie sub-6 GHz) \cite{3gpprel15}.
In the meantime, the O-RAN Alliance, a consortium of industry and academic partners, introduced important innovations for 5G Open-RAN architectures, among which, is the Open-RAN Intelligent Controller (RIC). This new architectural component provides a centralized abstraction of the RAN and facilitates custom control plane functions, \eg enabling closed-loop control via action and feedback loops between RAN components and their controllers\cite{oran}. 
\emph{In this work, we exploit the architectural advantage of Open-RAN architectures and the practicality of using directional transmitters on a UAV to design a closed-loop control system that jointly optimizes the locations and transmission directionality of the UAV to satisfy its application requirements without penalizing the neighboring ground users.}
Specifically, we employ wide-angle directional transmitters at the drone in uplink, e.g., a small directional antenna or a small phased array \cite{amorim2018measured, lyu2017blocking, BertizzoloHotmobile20}. With respect to bulky and power-hungry highly-directional transmitters, e.g., large directional antennas or large phased arrays, this solution is more suitable for small form-factor, battery-powered UAVs.
Moreover, this design choice is conservative with respect to the cellular paradigm which does not support high-directionality in FR1 and is compliant with the 5G NR standard which specifies a maximum of 4 antenna elements in uplink in FR1 \cite{3gpprel15}.
Finally, we envision an application scenario where the UAV is interested in capturing and streaming the live footage of a particular event or object located at a Point of Interest (POI). 
The drone would contact the control service running at the cellular infrastructure for instructions regarding the optimal aerial location and transmission directionality that satisfy the application's requirements.

In this article, we make the following contributions:
\newline $\bullet$ To motivate this work, we quantify the ground UEs' uplink throughput degradation (as much as $20\%$), in the presence of a drone on a dedicated full-stack cellular testbed. Moreover, we verify how controlling the drones' location and transmission directionality can improve the overall network performance (up to $28\%$).
\newline $\bullet$ We mathematically formulate the network control problem of a drone streaming video from a POI. Due to difficulty for the drone to physically relocate and adapt to the fast-changing dynamics of some of the involved network variables (e.g., radio resource scheduling and modulation schemes, which vary at the timescale of milliseconds), we reformulate the original problem using long-term average network information such as cell load and average uplink signal powers.
Then, we propose a control system solution for 5G Open-RAN architectures specifically to support drone-sourced video streaming. Our system features \textit{low overhead}, in-band communication (which makes it \textit{5G compatible}) and operates on a drone-initiated request-based approach that guarantees \textit{users' privacy}.
\newline $\bullet$ We prototype the proposed control solution in a real outdoor, multi-BS RAN cellular testbed. To the best of our knowledge, this is the first full-stack experimental assessment of intelligent control for cellular-enabled UAVs. 
Further, we assess the performance of our solution at scale through extensive simulations using the actual BSs topology and cell load profiles from a major US cellular operator for various scenarios. 
Our experiments demonstrate that our Open-RAN-based control can guarantee the UAV's UL QoS to steam an HD video while achieving \textit{effectiveness} (an average $20\%$ network capacity gain), \textit{adaptability} to changing network conditions, and \textit{scalability} to large cellular deployments.

The remainder of the paper is organized as follows. Section~\ref{sec:motivation} motivates the need for a solution to the drone-sourced uplink inter-cell interference through real-world measurements. Section~\ref{sec:formulation} formulates the drone-specific network control problem we aim to solve. Section \ref{sec:controlsystem} proposes the design of a closed-loop control system for Open-RAN architectures. The proposed approach is assessed through real-world experiments and extensive simulations in Section \ref{sec:evaluation}. We summarize the related work in Section~\ref{sec:relatedwork} and draw the main conclusions in Section ~\ref{sec:conclusion}.

\section{Motivational Experiments}
\vspace{-0.5mm}
\label{sec:motivation}
To motivate the need for a UAV-specific control solution, we perform a first-of-its-kind measurement campaign on a dedicated outdoor full-stack LTE testbed with a mixed set of aerial and ground UEs. Specifically, we aim to assess the ground UEs' \textit{uplink} performance degradation caused by one UAV connected to a different cell, as we vary the UAV's location, antenna pattern, and transmission directionality.
Our LTE-based analysis adopts the same low- and mid-band frequencies designated for drone cellular links in 5G NR (FR1).

\begin{figure*}[t!]
  \centering
  \subcaptionbox{
  Changing location: Deployment. 
  \label{fig:changeLocationDeployment}
  }
  {
    \includegraphics[width=.62\columnwidth]
    {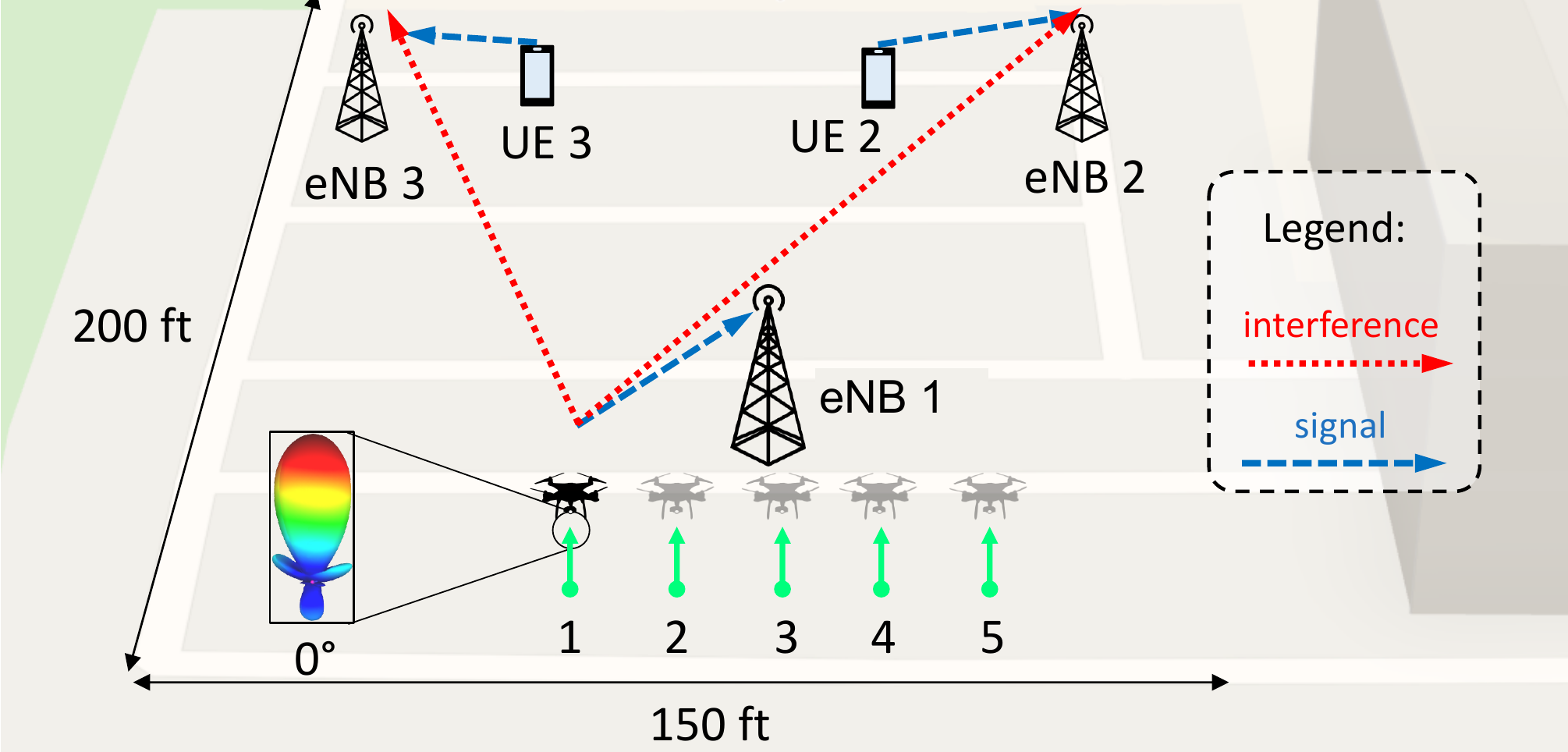}
  }
  \subcaptionbox{
    Ch. loc.: Performance. 
    \label{fig:changeLocationMeasure}
   }
  {
    \includegraphics[width=.34\columnwidth]
    {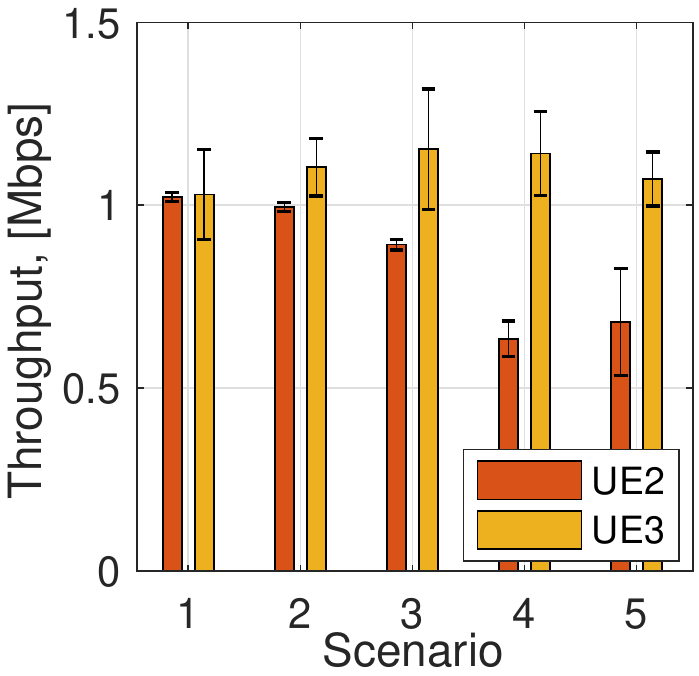}
  }
  \subcaptionbox{
  Changing directionality: Deployment.   
  \label{fig:changeRotationDeployment}
  }
    {
    \includegraphics[width=.62\columnwidth]
    {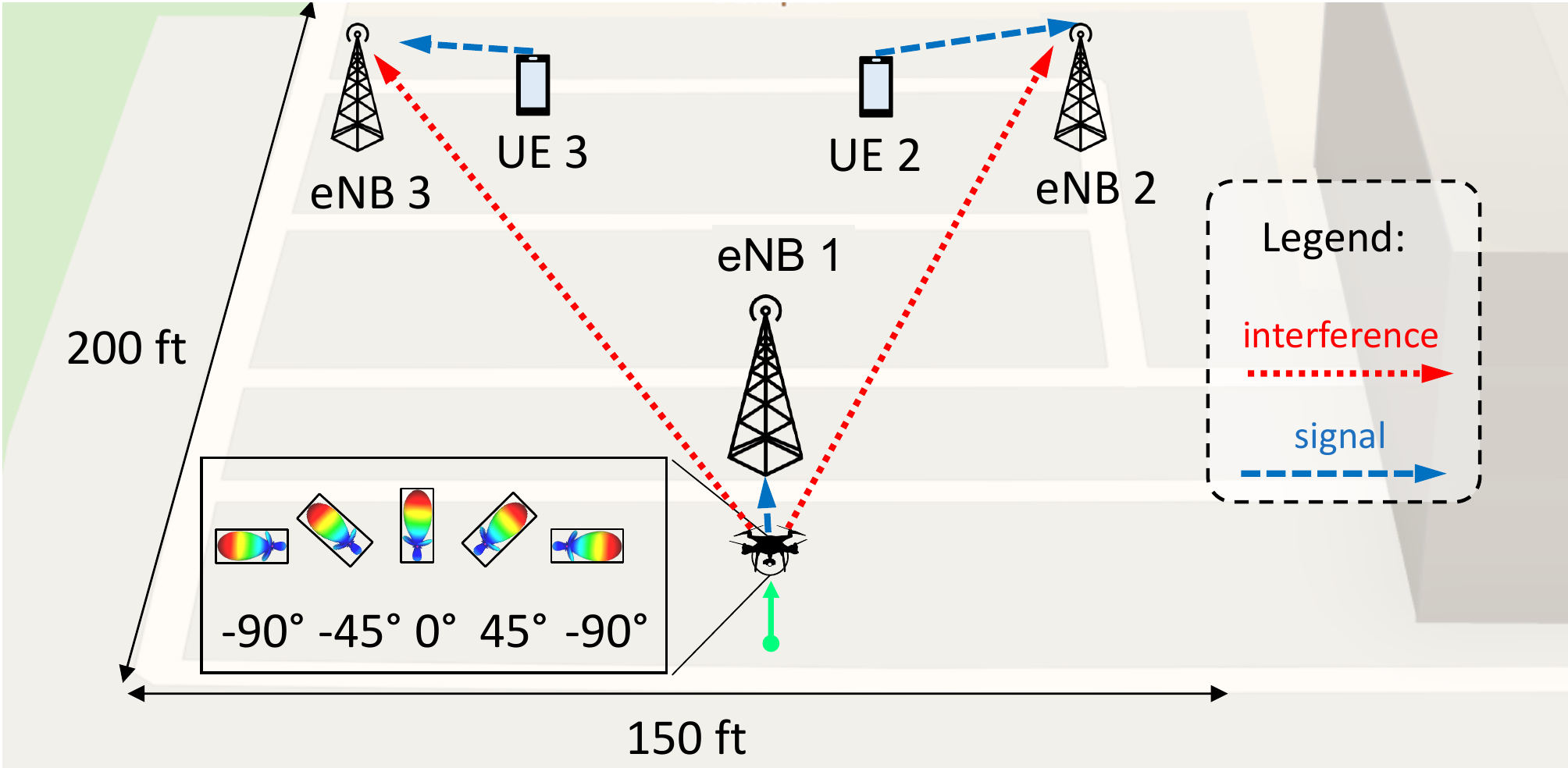}
  }
  \subcaptionbox{
    Ch. dir.: Performance. 
  \label{fig:changeRotationMeasure}
  }
  {
    \includegraphics[width=.34\columnwidth]
    {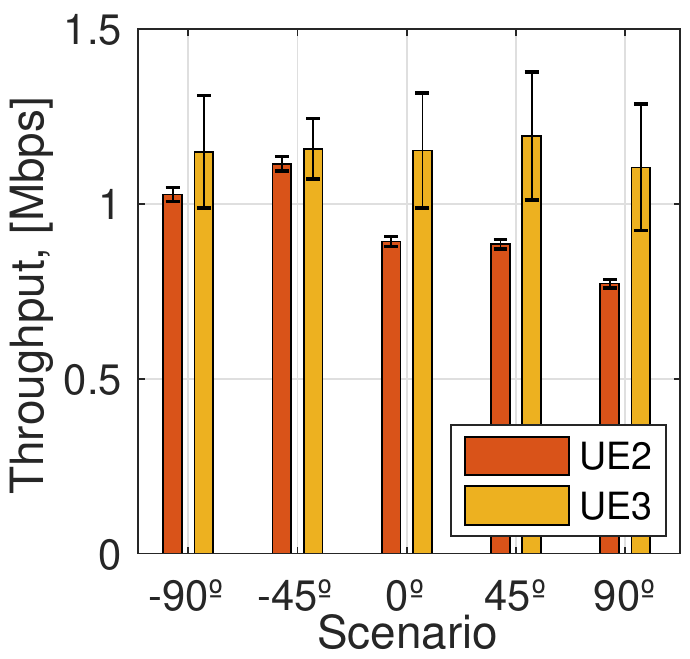} 
  }
  \vspace{0mm}
  \caption{Measurements for different UAV locations (and same TX directionality) and different TX directionality (and same location).}  
  \vspace{0mm}
  \label{fig:Intuition}
\end{figure*}

\vspace{0mm}
\subsection{UAV- and Terrestrial-UEs Coexistence}
\vspace{0mm}
Our testbed consists of three cells (eNBs), each serving one cellular user (UE) as illustrated in Fig. \ref{fig:motivation_topology}.
We refer the reader to Section \ref{sec:experiments} for a detailed description of the experimental setup, including the software, hardware, and testing methodology. 
Here, we analyze three scenarios:
\ul{Ground UAV}: the UAV acts as a traditional ground UE with an obstacle (a big van) positioned to mimic the typical blockage  experienced by pedestrians (Fig.~\ref{fig:motivationNLos}). The drone is equipped with an omnidirectional TX, and it should be considered as a traditional ground user for this case.
\ul{Omnidirectional TX}: The UAV-based UE is an aerial user flying higher than the obstacle height, employing a traditional omnidirectional antenna (Fig. \ref{fig:motivationLos}).
\ul{Directional TX}: The UAV-based UE is an aerial user flying higher than the obstacle height, employing a wide-angle directional antenna pointing at its serving BS (Fig.~\ref{fig:motivationLosDir}).

First, we assess the performance of the overall network in the absence of aerial links, that is when all the UEs are on the ground (Ground UAV). 
In this case, we do not fly the drone and keep it behind the positioned obstacle (see Fig~\ref{fig:motivationNLos}).
For each experiment in this section, we repeat a $10\:\mathrm{second}$-long measurement for $20$ times and record the per-user uplink throughput and its standard deviation in Megabits per second ($\mathrm{Mbps}$). The network performance for the first experiment is reported in Fig.~\ref{fig:motivationPerformance} (`Ground'). To better quantify the interference dynamics, we  also measure the single-user upper-bound performance measured in isolation, that is, when no other users are present in the network. We report this as the red dashed line in Fig.~\ref{fig:motivationPerformance}.
When all UEs are on the ground, the average per-user uplink throughput is $2.1\:\mathrm{Mbps}$, while the aggregate network throughput is $6.3\:\mathrm{Mbps}$.
Next, we assess the performance of our  network when flying the UAV-based user.
In this scenario, the UAV-based user hovers above the obstacle height, and thus benefits from the LoS condition with multiple eNBs as shown in Fig.~\ref{fig:motivationLos}. 
We measure the network performance and report them in Fig.~\ref{fig:motivationPerformance} (`Omni. TX'). 
The aerial LoS favors the UAV which records $1.76\:\mathrm{Mbps}$ ($+65\%$ over the `Ground' scenario), however, the performance of UE 2 and UE 3 drop to $2.61\:\mathrm{Mbps} (-12\%)$ and $1.55\:\mathrm{Mbps} (-31\%)$, respectively. 
This scenario degrades the performance of neighboring cells by $20\%$ compared to  the `Ground' scenario.

In our third deployment, the UAV flying above the obstacle height is equipped with a steerable directional transmitter (directional antenna) pointed toward the dominant LOS path with the serving eNB (see Fig.~\ref{fig:motivationLosDir}).
In this case, the average per-user uplink throughput of $2.2\:\mathrm{Mbps}$ and the aggregate network performance of $7.6\:\mathrm{Mbps}$ as reported in Fig.~\ref{fig:motivationPerformance} to the right (`Dir. TX'). Equipping the UAV with a directional transmitter benefits the UAV itself ($+25\%$ throughput as compared to the `Omnidirectional TX'  scenario) and the other UEs as well ($+11\%$ and $+60\%$, respectively).
Overall, employing a directional transmitter at the UAV increases the UAV data-rate by $28\%$ and reduces the inter-cell interference by $30\%$ with respect to the `Omnidirectional TX' scenario.

\subsection{Effect of UAV Parameters}
Here, we measure and observe how the performance of the prototyped cellular network varies with the UAV's location and its transmission directionality.
We treat the two cases in isolation.

\textbf{Changing the UAV's Location.}
Figs.~\ref{fig:changeLocationDeployment} and ~\ref{fig:changeLocationMeasure} report the deployment and performance of the 3-eNB RAN network for the UAV at $5$ different locations, connected to eNB $1$. As the UAV moves to $5$ different locations, \ue{2} and \ue{3} experienced uplink rate variation of up to $30\%$. This experiment suggests that the UAV location can be controlled to tune the UAV impact on the rates of the neighboring ground users.


\textbf{Changing the UAV's TX Directionality.}
Figs.~\ref{fig:changeRotationDeployment} and \ref{fig:changeRotationMeasure} show the deployment and performance as we fix the UAV's location and vary its transmission directions, namely $-90^{\circ}, -45^{\circ}, 0^{\circ}, 45^{\circ}$, and $90^{\circ}$ with respect to its LOS direction with its serving eNB.
The $5$ UAV transmission directions correspond to different \ue{2} and \ue{3} rates, with variations up to $38\%$. This experiment suggests that the UAV transmission directionality can be controlled to tune the UAV impact on the neighboring ground users' rate.

\smallskip
To summarize, we have verified the considerable implications of extending cellular support to UAVs through real-world experiments on a dedicated multi-cell RAN testbed (up to $20\%$ throughput degradation to neighboring ground UEs). At the same time, we have seen that changing the location in space of the UAV equipped with a directional transmitter and changing the UAV's TX directionality can positively impact the performance of ground UEs in the surrounding (up to $38\%$ in uplink).
In the next sections, we will study how to intelligently control these parameters to minimize the impact of the UAV uplink transmissions on neighboring ground users and improve the overall uplink network performance.

\section{Connected-Drone Control Problem}
\label{sec:formulation}
\glsresetall
In this section, we formally define and mathematically formulate the specific Connected-Drone Control Problem (CDCP) of a drone intending to stream video content from nearby a Point of Interest (POI). Then, we iterate over the formulation and simplify its parameters to be observable and easily retrievable by an ISP. Finally, we explain how to solve the newly formulated problem via convex optimization.

\subsection{CDCP Formulation}
\begin{figure}[t!]
  \centering
  \subcaptionbox{
  CDCP Deployment.
  \label{fig:scenario}
  }
  {
    \includegraphics[width=.63\columnwidth]
    {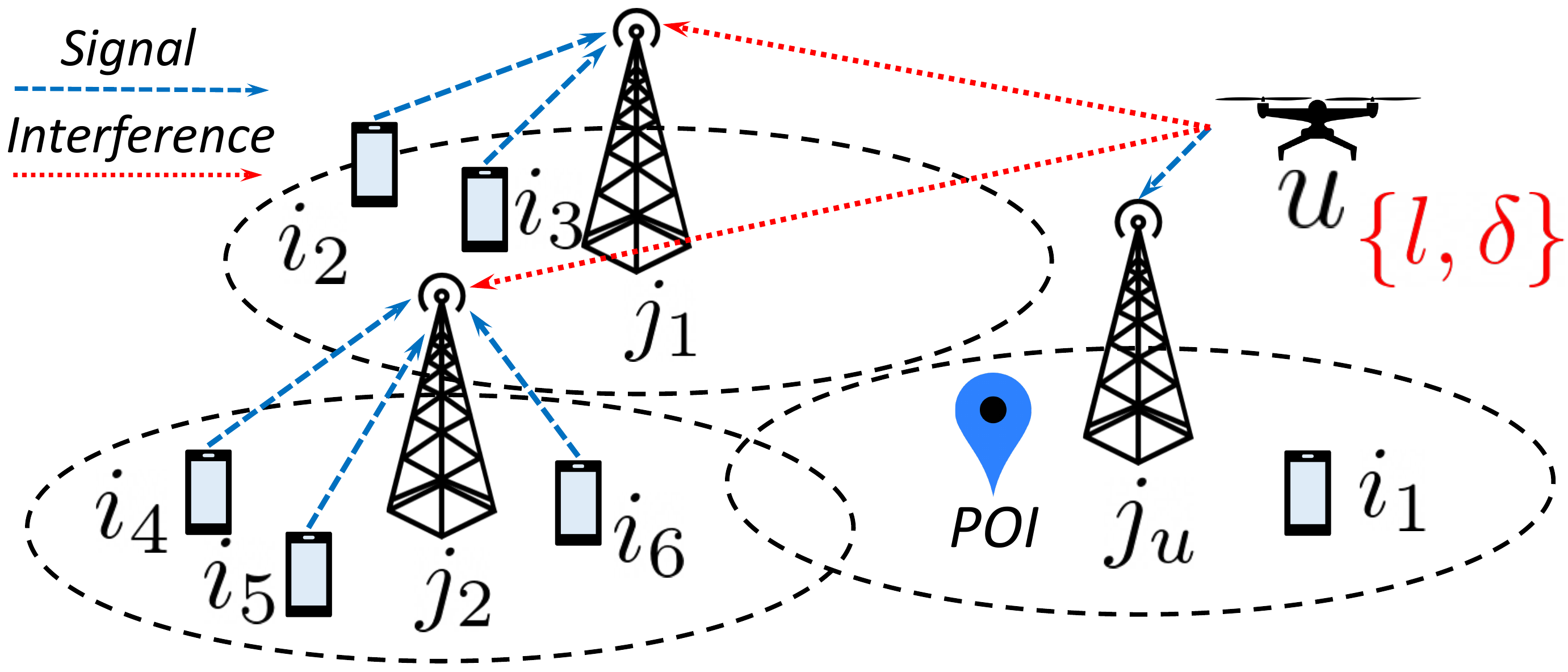}
  }
  \subcaptionbox{
    CDCP Interference. 
    \label{fig:uplinkinterference}
  }
  {
    \includegraphics[width=.3\columnwidth]
    {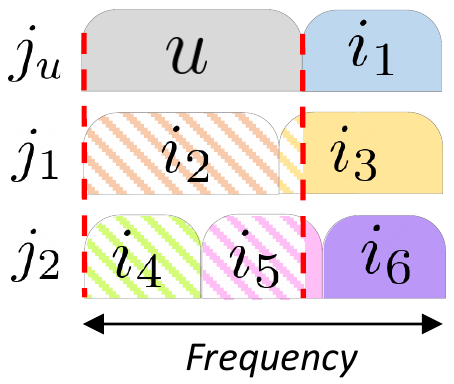}
  }
  \vspace{0mm}
  \caption{Connected-Drone Control Problem (CDCP) scenario.} 
  \label{fig:ncp}
\end{figure}

According to the 3GPP's directives which suggest cellular UAV operations in the 5G NR low- and mid-band (\ie sub-\ghz{6}), we herein consider the signal propagation characteristics of sub-\ghz{6} spectrum \cite{3gpprel15}.
We consider a small region in a traditional cellular network deployment where a set of \glspl{bs} $\mJ$, serve multiple ground cellular users denoted by the set $\mI$. Specifically, each BS $j \in \mJ$ serves a finite number of users denoted by the set $\mathcal{I}_j$, where the different sets $\mathcal{I}_j$ are disjoint, \ie $\;  \mathcal{I}_j \cap  \mathcal{I}_i = \emptyset \; \forall i \neq j$. 
Let us denote the number of UEs served by \gls{bs} $j$ as $w_j = |\mathcal{I}_j|$.
Additionally, there is a UAV-based user, denoted by $u$ that is served by \gls{bs} $j_u \in \mJ$ as illustrated in Fig.~\ref{fig:scenario}.  Accordingly, $w_u = |\mathcal{I}_u| + 1$ when the UAV is present.
At any given time, a cellular UE is allocated a finite number of uplink and downlink \glspl{prb} by the 5G \gls{bs} scheduler. 
At a given time instant, let us indicate with $x_i$ the number of \textit{uplink} \glspl{prb} allocated to user $i$.
Without loss of generality, we assume all the cells operating on the same carrier frequency with bandwidth $B$, and each user $i$ is allocated a bandwidth $B_i$ that equals $x_i$ times the bandwidth of a \gls{prb} $B_{PRB}$. 
Hereafter, all traffic analysis implicitly refers to uplink traffic only.

As per the 5G \gls{nr} physical layer specification, cellular users' uplink channel access happens in \gls{cpofdm} or \gls{dftsofdm}. In this specification, when the cellular users access channels with overlapping frequencies, they interfere with each other as reported in Fig.~\ref{fig:uplinkinterference}. 
As we discussed and experimentally demonstrated in the `Ground' scenario in Section \ref{sec:motivation}, the impact of ground UEs-generated interference on other ground users is marginal as their uplink power to neighboring cells is attenuated by surrounding obstacles such as buildings, vehicles, or trees.
\textit{It is instead a more serious issue when the interference comes from drones up in the sky. }

Let us define $x_{iu}$ as the number of uplink \glspl{prb} that user $i$ served by \gls{bs} $j_i$ is co-scheduled with those of the UAV served by \gls{bs} $j_u$.
The $x_{iu}$ parameters are integer numbers that equal zero when the ground user $i$ has no shared resource allocation with the UAV $u$, \eg $x_{iu} = 0$ for $i = 1, 6$ 
, and are non-zero positive integers otherwise, \eg  $x_{iu} > 0$ for $i = 2, 3, 4, 5$ (see Fig. \ref{fig:uplinkinterference}).
Due to the orthogonality of intra-cell scheduling decisions, it is guaranteed that $x_{iu} = 0$ for all the users that are in the same cell $j_u$ serving the UAV, \eg for $i = 1$, while $x_{iu} > 0$ for at least one user $i$ served by the cell $j_i \neq j_u$, that is, $\sum_{i \in \mI_j} x_{iu} > 0 \; \forall j \neq j_u \in \mJ$.
Last, we normalize the coefficients $x_{iu}$ as $\hat{x}_{iu} = x_{iu}/x_{i}$. The coefficients  $\hat{x}_{iu} \in [0, 1]$ represent the band share that user $i$ has with the UAV $u$. Conversely, $\hat{x}_{ui} = x_{iu}/x_{u}$ represent the UAV band share with the ground user $i$.
Let us define the uplink signal power of user $i$ received at the serving \gls{bs} $j_i$ as $P^{i}$. Similarly, we define the UAV received signal power at its serving \gls{bs} $j_u$ as $P^{u}(l, \delta)$, function of the UAV's location $l$ and the UAV's serving BS's location $l_u$ (with $l, l_{j_u} \in \mathcal{L}$, where $\mathcal{L}$ is the set of possible 3D locations in space) and of the UAV's 3D transmission direction $\delta = (\delta_a, \delta_e) \in \{[0, 2\pi], [0, \pi]\}$ (where $\delta_a$ and $\delta_e$ are the azimuth and elevation directions). 
Similarly, $P_{j}^{u}(l, \delta)$ denotes the UAV's uplink interference power at cell $j \neq j_u$ as a function of $l, l_j$ and $\delta = (\delta_a, \delta_e)$.
The received UAV's power at a BS j is calculated through 
\begin{equation}\small
    P_{j}^{u}(l, \delta) = P^u_{\text{TX}} - L_{\text{PL}}(l,l_j) + G_{\text{Dir}}(\delta, l_j)
\end{equation}
where $P^u_{\text{TX}}$ is the UAV transmitted power (maximum \dbm{23} for 5G UEs), $G_{\text{Dir}}(\delta, l_j)$ is the angular gain of the employed directional transmitter with respect to the BS's $j$ location $l_j$, and  $L_{\text{PL}}(l,l_j)$ is the uplink Air-to-Ground (A2G) path-loss attenuation among the two locations $l$ and $l_j$. The latter is inclusive of effects such as BS antenna's patterns, vertical coverage, and cells' down-tilting as well as height-dependent losses, cell sectorization, and drone fluctuations. Here, we report the formulation described in \cite{amorim2017radio}
\begin{equation}\small
     L_{\text{PL}}(l, l_j) = \alpha 10 \log_{10} (d) + \beta + X_{\sigma}
\end{equation}
where $d$ is the distance between the UAV and the \gls{bs} location $l_j$, while $\alpha$, $\beta$, and $\sigma$ are the UAV's height-dependent path-loss exponent, the intercept point with the $d = $ \meter{1} line, and the standard deviation of the normally distributed shadowing $X$, respectively. The reader is referred to \cite{amorim2017radio} for further details. 
Last, let $N$ be the \gls{npsd}. 
Accordingly, the uplink \gls{sinr} for user $i$ served by cell $j$ is denoted by
$ \mathrm{SINR}_{i} = P^{i} / (N B_{i} + P^{u}_{j} \hat{x}_{ui}) $. As we consider the uplink interference from ground UEs to the UAV negligible, the UAV's \gls{sinr} can be calculated as 
$ \mathrm{SINR}_{u} = P^{u} / N B_{u} $.
Consequently, the uplink achievable rate $r_i$ of each UE $i$ (expressed as $r_u$ for the UAV) is upper bounded by the Shannon's capacity
$C_i = B_i log_{2}(1+{\mathrm{SINR}}_{i})$
and is a function of the selected physical-layer coding and modulation scheme $m_{i}$ \cite{mogensen2007lte}.
In Section \ref{sec:introduction}, we illustrated a series of applications that rely on the availability of a cellular-connected UAV to stream a live video of a specific Point of Interest (POI).
It is reasonable to think, thus, that one of the drone's application requirements is to be within the coverage area of the POI. 
Being POI a 3D location in space $\hat{l} \in \mathcal{L}$, we express this application-specific constraint in the form of respecting a maximum distance $\text{dis}_{\text{max}}$ from the POI
Last, we assume that the drone serving BS $j_u$ is selected upon a given policy by the ISP, as it happens for terrestrial users, and it is known at any given time.

From a network optimization perspective, adopting the drone's location $l$ and its transmission directionality $\delta$ as network control variables, the cellular CDCP can be formulated as follows:
\begin{align}\small
&\underset{\substack{l, \delta} }{\text{maximize}}  \; \;
r_u + \sum_{j \in \mJ / j_u}
\sum_{i \in \mI_j} \: 
r_i \;,
\label{eq:max}\\
& 
\text{subject to} 
\hspace{0.2cm} 
r_{i} > r_\text{min} \;, \hspace{2cm} \forall i \in \mI
\label{eq:con:min_rate} \\
& 
\hspace{1.5cm} 
r_{u} > r_\text{app}\;,  
\label{eq:con:min_video}
\\
& 
\hspace{1.5cm} 
|l - \hat{l} | < \text{dis}_{\text{max}}.
\label{eq:con:distance}
\end{align} 
The CDCP problem formulated in (\ref{eq:max})-(\ref{eq:con:distance}) aims at maximizing the aggregate cellular users' uplink throughput. Specifically, the objective of CDCP is to maximizes the total uplink data rates of all users that are co-scheduled (i.e., having an overlapping frequency band) with the UAV, including the UAV itself. Optimizing throughput for other users is outside the scope of this work.
By maximizing the aggregate users' rate, the CDCP in (\ref{eq:max})-(\ref{eq:con:distance}) is robust against solutions that over-penalize a few users for the sake of overall uplink network throughput.
The rationale behind this intrinsic design choice is two-fold.
First, in the limited-band regime, limited-band users are more sensitive to performance drops caused by additional UAV interference due to their already limited data-rate. Second, in commercial cellular deployments, limited-band users are generally grouped in highly populated cells and usually outnumber wide-band users that are instead served by under-populated cells. 
Accordingly, by solving the CDCP for the maximum uplink network throughput guarantees reduced interference share toward small-bandwidth users (which are the majority and are the more sensitive toward throughout drops) while most of the UAV's interference share is handled by a few ground users that benefit from large bandwidth chunks and are less sensitive to data-rate drops.
This control solution thus intrinsically provides fairness to the system by protecting the majority of the carrier's paying subscribers, which populate dense cells and are less robust to data-rate drops.

As an example, referring to Fig. \ref{fig:uplinkinterference}, the CDCP maximizes uplink rates of users $i_2$ and $i_4$, which share their whole bandwidth with the interfering UAV, \ie $\hat{x}_{iu} = 1$, together with the partially overlapping rate of users $i_3$ and $i_5$ ($0 < \hat{x}_{iu} < 1$ ). On the other hand, the rate of user $i_6$, for which $\hat{x}_{iu} = 0$, is factored out of the CDCP, whereas rate of user $i_1$ is a constant of the CDCP and has been excluded from the formulation for this reason.
Moreover, the CDCP prioritizes limited-band users, \eg user $i_4$ and $i_5$, over large-band users such as $i_2$ and $i_3$.
Together with the maximization function (\ref{eq:max}), we expressed constraints regarding the minimum per-user uplink QoS (\ref{eq:con:min_rate}), the minimum UAV application-specific uplink data-rate (\ref{eq:con:min_video}), and the maximum distance with respect to a given POI  (\ref{eq:con:distance}).
In other words, by selecting the optimal drone's location and its transmission directionality, the CDCP (\ref{eq:max})-(\ref{eq:con:distance}) aims at minimizing the UAV's impact on bandwidth-sensitive ground UEs, while guaranteeing application-layer requirements to the drone.

\smallskip
\textit{The CDCP formulated in (\ref{eq:max})-(\ref{eq:con:distance}) is, however, hard to solve for the following reasons:} \newline \indent 
\begin{enumerate*}[label=(\roman*)] 
\item The time-varying, fine-grained information about $x_i$ and $m_i$ are determined by the BSs' MAC schedulers and PHY implementation which are often not exposed to a drone controller residing outside those BSs, due to transport overhead and/or proprietary PHY/MAC implementations~\cite{larsen2018survey}. 
\newline \indent\item Even though the latest development in the industry is pushing some PHY/MAC BS's control logic to be exposed to the northbound controller (e.g., the Open-RAN Intelligence Controller~\cite{oran}), the scheduling and rate-adaptation decisions are re-calculated in every transmission interval (e.g., $1~\rm{ms}$). It is therefore unpractical for a solver to use this information in time, and for the drone to physically relocate to a new location in space with the instantaneous variations of $x_i$ and $m_i$. As a result, the coefficients $x_i$ and the modulations $m_i$ in (\ref{eq:max})-(\ref{eq:con:distance}) are not only outside the control of the drone controller, but are non-observable.
\newline \indent \item In a real cellular network, ground users frequently relocate. Their mobility, even within the very same cell, changes their uplink signal strength at the serving \gls{bs} at a fast pace. This makes their received power $P^{i}$ and thus their uplink rate $r_i$  in (\ref{eq:max})-(\ref{eq:con:distance}) change in real-time. 
Under this condition, the solver might calculate optimal solutions that, when implemented, do not reflect the current network state. 
\end{enumerate*}

\noindent
\textbf{CDCP Reformulation}:
\textit{To address these challenges, we reformulate the CDCP in (\ref{eq:max})-(\ref{eq:con:distance}) by employing information that is changing at a relatively slower pace and is feasibly retrievable at the \gls{isp} in an Open-RAN architecture} \cite{oran}. Specifically, we explore the network parameters that are described in the following. 
Let us first define a time period $\mathcal{T}$. The following variables are long-run averages defined over this time interval.

Let $\widetilde{w}_j$ be the average number of users at a cell $j$ (also known as cell's load).
On an Open-RAN architecture, this information can be retrieved at intervals as short as $1~\rm{second}$.
Then, we assume that under a proportional fairness scheduler, each of the $\widetilde{w}_j$ users served by BS $j$ will be allocated average bandwidth resources equal to $\widetilde{B}_j = B / \widetilde{w}_j$.
Further, let $\widetilde{P}_j$ be the aggregate uplink signal power received at \gls{bs} $j$ over band $B$ in absence of UAVs. This information can be calculated using statistical users' distribution and the average cell load $\widetilde{w}_j$, that is, the average number of RRC connected UEs.
Accordingly, the average UAV's uplink \gls{sinr} at its serving \gls{bs} $j_u$ can be expressed as ${\widetilde{\mathrm{SINR}}}_u(l, \delta) = \frac{P^{u}(l, \delta)}{N\widetilde{B}_{u}}$, 
where $P^{u}(l, \delta)$ is the received UAV signal power at the associated \gls{bs} $j_u$. The latter can be evaluated from the UAV location $l$ and its transmission directionality $\delta$. 
The average \gls{sinr} for other users $i \in \mI_u$, can be defined as ${\widetilde{\mathrm{SINR}}}_{i} = \frac{\widetilde{P}_u}{NB}$ and is considered a constant of the CDCP.
Similarly, the average uplink \gls{sinr} for ground users served by \gls{bs} $j \neq j_u$ is defined as ${\widetilde{\mathrm{SINR}}}_j(l, \delta) = \frac{\widetilde{P}_j}{NB + P^{u}_{j}(l, \delta)}$ where $P^{u}_{j}(l, \delta)$ is the received UAV interference power at the \gls{bs} $j \neq j_u$, which is a function of the UAV's location $l$ and transmission directionality $\delta$ relative to the location of BS $j$. 
The CDCP (\ref{eq:max})-(\ref{eq:con:distance}) can thus be reformulated as follows:
\begin{align} \small
\label{eq:maxsinr}
\underset{\substack{l, \delta }
}{\text{maximize}}  \;\; \small
&B_u \log_2 \big(1 +\widetilde{\mathrm{SINR}}_{u}(l, \delta) \big) \;\; + \\  \small
\nonumber &\sum_{j \in \mJ / j_u}  B \log_2 \big(1 + \widetilde{\mathrm{SINR}}_{j}(l, \delta) \big) ,\\ \small
\text{subject to} 
\hspace{0.2cm} 
&\widetilde{\mathrm{SINR}}_{j}(l, \delta) > \mathrm{SINR}_\text{min} \quad\quad\quad\quad \forall j \in \mJ / j_u
\label{eq:con:min_sinr}, \\ \small
\hspace{1.5cm} 
&\widetilde{\mathrm{SINR}}_{u}(l, \delta) > \mathrm{SINR}_\text{app}  
\label{eq:con:min_video_sinr},
\\ \small
\hspace{1.5cm} 
&|l - \hat{l} | < \text{dis}_{\text{max}}, 
\label{eq:con:distance_sinr}
\end{align} 
where (\ref{eq:maxsinr}) is the statistical average over $\mathcal{T}$ of (\ref{eq:max}) and  (\ref{eq:con:min_sinr})-(\ref{eq:con:min_video_sinr}) are the statistical averages of constraints (\ref{eq:con:min_rate})-(\ref{eq:con:min_video}) over $\mathcal{T}$.

Different from the variables involved in (\ref{eq:con:min_rate})-(\ref{eq:con:min_video}), the parameters $P^{u}$, $P^{u}_{j}$, $\widetilde{P}_j$, $\widetilde{w}_j$, and $\widetilde{B}_{u}$ employed in our reformulation are statistical averages that help us avoid the dependency on the non-stationary instantaneous terms. 
Moreover, all the parameters employed in our reformulation are fully observable. Indeed, $P^{u}$ and $P^{u}_{j}$ can be calculated knowing the drone's location $l$ and its transmission directionality $\delta$, while the average received uplink powers $\widetilde{P}_{j}$, cell loads $\widetilde{w}_j$, and the average users' band occupancy $\widetilde{B}_{u}$ can be retrieved through the \glspl{bs}'s control APIs available on Open-RAN architectures or calculated by the ISP using statistical UEs' distributions.

\textit{With $\mathcal{T}$ correctly dimensioned, the CDCP reformulation (\ref{eq:maxsinr})-(\ref{eq:con:distance_sinr}) features observable and easily retrievable network parameters.}
Specifically, $\mathcal{T}$ must match three constraints and should be dimensioned accordingly. $\mathcal{T}$ must be larger than the latency to retrieve information from the neighboring BSs, in the order of $1~\rm{second}$ in modern Open-RAN architectures \cite{oran}. $\mathcal{T}$ must also be larger than the mathematical solver run-time employed to solve the optimization problem. This last constraint depends on the hardware and software configuration used to execute the mathematical solver. We will see later that this parameter is in the order of seconds in our prototype and is expected to be in the order of $1~\rm{second}$ for commercial hardware and software configurations. 
Last, $\mathcal{T}$ must be dimensioned to catch substantial users' macro mobility patterns throughout the day, \eg the home to office commute, or traffic load swifts between cells due to events and gatherings. This last constraint is deployment-dependent and should be evaluated carefully, depending on the traffic load dynamics. In most cases, $\mathcal{T} \leq 10$ minutes is a satisfactory range. 
\textit{With $\mathcal{T}$ correctly dimensioned, for example in the order of minutes, the CDCP (\ref{eq:maxsinr})-(\ref{eq:con:distance_sinr}) can always be solved in a timely fashion and the solution can be effectively implemented by an \gls{isp}.
}
Finally, at every interval $\mathcal{T}$, the problem is to be solved once again for a new optimal solution.


\textbf{Solving the CDCP.} 
From an optimization standpoint, the CDCP (\ref{eq:maxsinr})-(\ref{eq:con:distance_sinr}) features five control variables (the three dimensions of the UAV location and the two dimensions of the TX direction) and an optimization function with many local minima valleys separated by large barriers (without loss of generality we negate (\ref{eq:maxsinr}) to solve the CDCP via convex optimization). 
Here, for local minimum we intend a control variables' combination whose function is lower than a set of solution points around it, and for valley, the set of these solution points separated by the large barriers.
The intuition that we leverage to solve the CDCP with a bounded number of parallel solvers is that, in a physical network deployment, the high barriers of the utility function (\ref{eq:maxsinr}) are for TX directions pointing at neighboring BSs and some UAV locations in the delimited area near the POI.
Accordingly, the valleys that contain the local minima solutions are for TX directions in between two neighboring BSs (and some initial UAV location in space). 
Accordingly, the number of minima of the function (\ref{eq:maxsinr}) is in the order of the number of interfered BSs in the surrounding: $\mathcal{O}(|\mathcal{J}|)$.
As wireless carriers carefully chose BSs sites to exploit spatial diversity and limit inter-cell-interference at any given band, we argue that the number of interfered BSs in the surrounding $|\mathcal{J}|$ is bounded and finite for any deployment. 
Most importantly, this number does not scale with the size or density of the deployment or frequency band in use. 
In fact, in commercial cellular networks, the UEs' output power is regulated by the 5G NR standard, while a more aggressive path loss compensates for the deployment density at higher frequencies. 
Consequently, the maximum ‘number of BSs interfered by the UAV' is bounded and constant at any 5G band and in any BSs deployment scenario. Accordingly, the number of local minima of our function is also bounded, which ensures the formulated CDCP (\ref{eq:maxsinr})-(\ref{eq:con:distance_sinr}) can always be solved in a timely fashion.
To practically solve the CDCP, we employ a global convex optimization solver of the kind basin-hopping. Specifically, knowing the approximate characterization of the solution space, we solve the CDCP by searching the valleys in the directions between interfered neighboring BSs. 
We use a pool of $|\mathcal{J}|$ convex optimization solvers that can run in parallel. 
When the solvers return the value of the found local optimal operational point, the global optimizer compares them and returns the global optimal solution to be implemented.

With peak cruise speeds of \metersec{20}, the UAV only takes a few seconds to relocate and implement the new solution.
Lastly, the proposed CDCP formulation is flexible enough to accommodate stringent streaming application requirements, \eg $\text{dis}_{\text{max}} = 0$, or to solve for the best achievable drone QoS should the CDCP become infeasible, by progressively relaxing constraint (\ref{eq:con:min_video_sinr}).
Next, we propose a control system architectural solution to practically solve CDCP (\ref{eq:maxsinr})-(\ref{eq:con:distance_sinr}) in real RANs.
\begin{figure}[t!]
\centering
\includegraphics[width=.95\columnwidth]{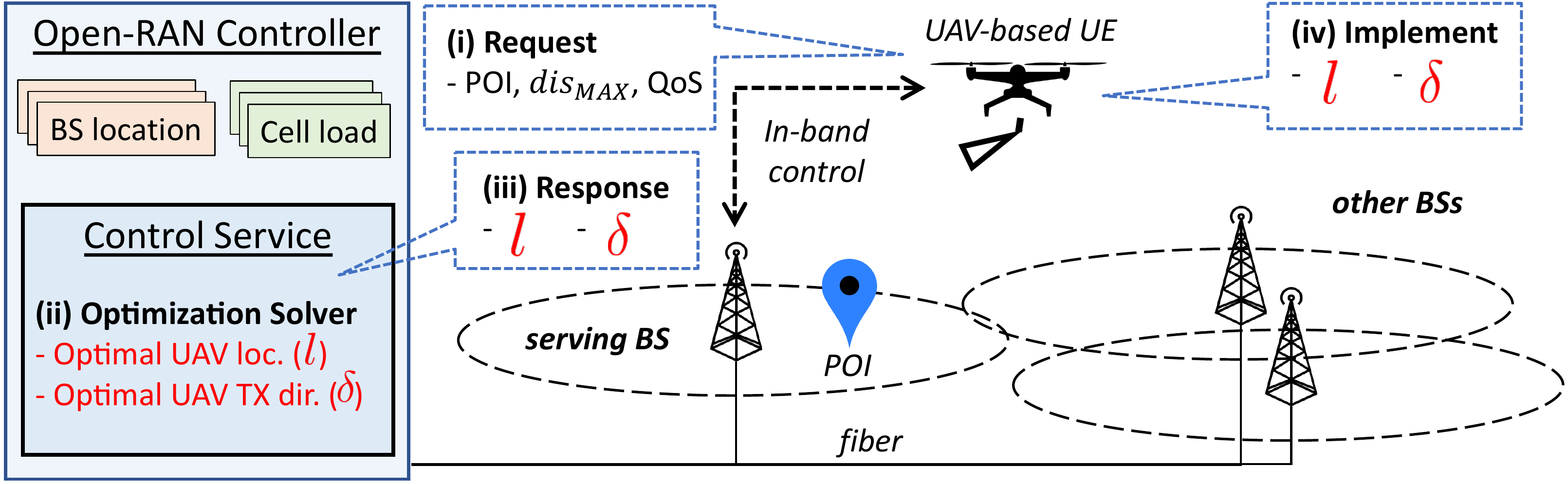} 
\caption{Proposed Open-RAN-based control system.}
\label{fig:controller}
\vspace{0mm}
\end{figure}

\section{Control System Design}
\label{sec:controlsystem}
To realize the optimal solution obtained by solving the CDCP above, we envision here a control service that is integrated with the cellular network infrastructure and that has access to both statistical and real-time information from multiple eNBs within a market region~\cite{kitindi2017wireless, gavrilovska2020cloud}. Some statistical and real-time information that is relevant for our drone control service is cell configuration (e.g., BS's location, antenna sector, and frequency band) and cell load (e.g., number of concurrent active users, averaged over every $1\rm{s}$ interval), respectively. Such control system is architecturally consistent with several efforts in the O-RAN Alliance ~\cite{oran} that envisages a Radio Access Network Intelligent Controller (RIC) platform deployed by the cellular operator that will host such services \cite{bonati2020open}. 

\textbf{Control System's Operations:} The envisioned control system, depicted in Fig.~\ref{fig:controller}, features a mathematical solver that is in charge of solving the CDCP formulated in Section \ref{sec:formulation} in near-real-time, that is, based on the current state of the network. Further, the service is in charge of relaying the computed optimal solution to the UAV over the same cellular links the drone is connected to, in an integrated-service-and-control fashion. 
The main operations of our control system are the following:\newline \indent
\begin{enumerate*}[label=(\roman*)]
\item The drone sends information to the control service about the POI's location, the application data-rate requirements, and the maximum tolerable distance from the POI ($\text{dis}_{\text{max}}$). 
\newline \indent \item Based on the drone's requirements, coupled with the network information (cell loads and topology), the Open-RAN control service computes the optimal solution (drone location, and drone's transmission directionality) by solving CDCP (\ref{eq:maxsinr})-(\ref{eq:con:distance_sinr}). 
\newline \indent\item The service sends back the computed solution (drone's best location, and best transmission directionality) to the drone. 
\newline \indent\item The drone autonomously implements the computed optimal solution that guarantees the application data-rate and location requirements while minimizing its impact on the ground users in the surrounding.
\newline \indent\item To cope with the changes in network load, the control service will periodically recalculate optimal solutions at every time interval $\mathcal{T}$, and communicate the new solution to the drone. At the same time, if there are any application requirement changes (e.g., new uplink QoS and/or a new POI), the drone will send these updates back to the control service which will directly proceed to a new calculation.
\end{enumerate*}

An illustration of our envisioned control loop architecture is shown in Fig. \ref{fig:controller}.
The proposed control service design is beneficial for many reasons. 
\newline \textit{- 5G Compatibility:} 
The control service at the Open-RAN controller interacts with the drone in-band, that is over the same cellular links that are used for UE connectivity (\eg the 5G data-link). This design choice eliminates the need for dedicated control links, Physical- or MAC layer modifications, and ultimately guarantees backward compatibility with the 5G NR standard stack.
\newline\textit{- Low overhead:}
Different from other interference mitigation schemes \cite{izydorczyk2018experimental, mei2019uplink}, our approach is extremely lightweight. The information retrieved by the solver and the response to the drone can be encapsulated in only a few bytes of information. This property is beneficial for two reasons. Having low-overhead guarantees fast solution calculation and prompt implementation at the drone. Also, light signaling preserves the RAN stack functionality, thus preventing channel saturation or undesirable delays.
\newline \textit{- Users' privacy:}
The proposed control approach is drone-initiated and guarantees aerial users' privacy. The drone voluntarily communicates its target POI and desired QoS to the service and implements the proposed solution autonomously. This way, no active trajectory control or motion tracking is performed by the ISP.

\begin{figure}[t!]
  \centering
  \subcaptionbox{
    UAV-based UE hardware schematics. 
  \label{fig:schematics}
  }
  {
    \includegraphics[width=.8\columnwidth]
    {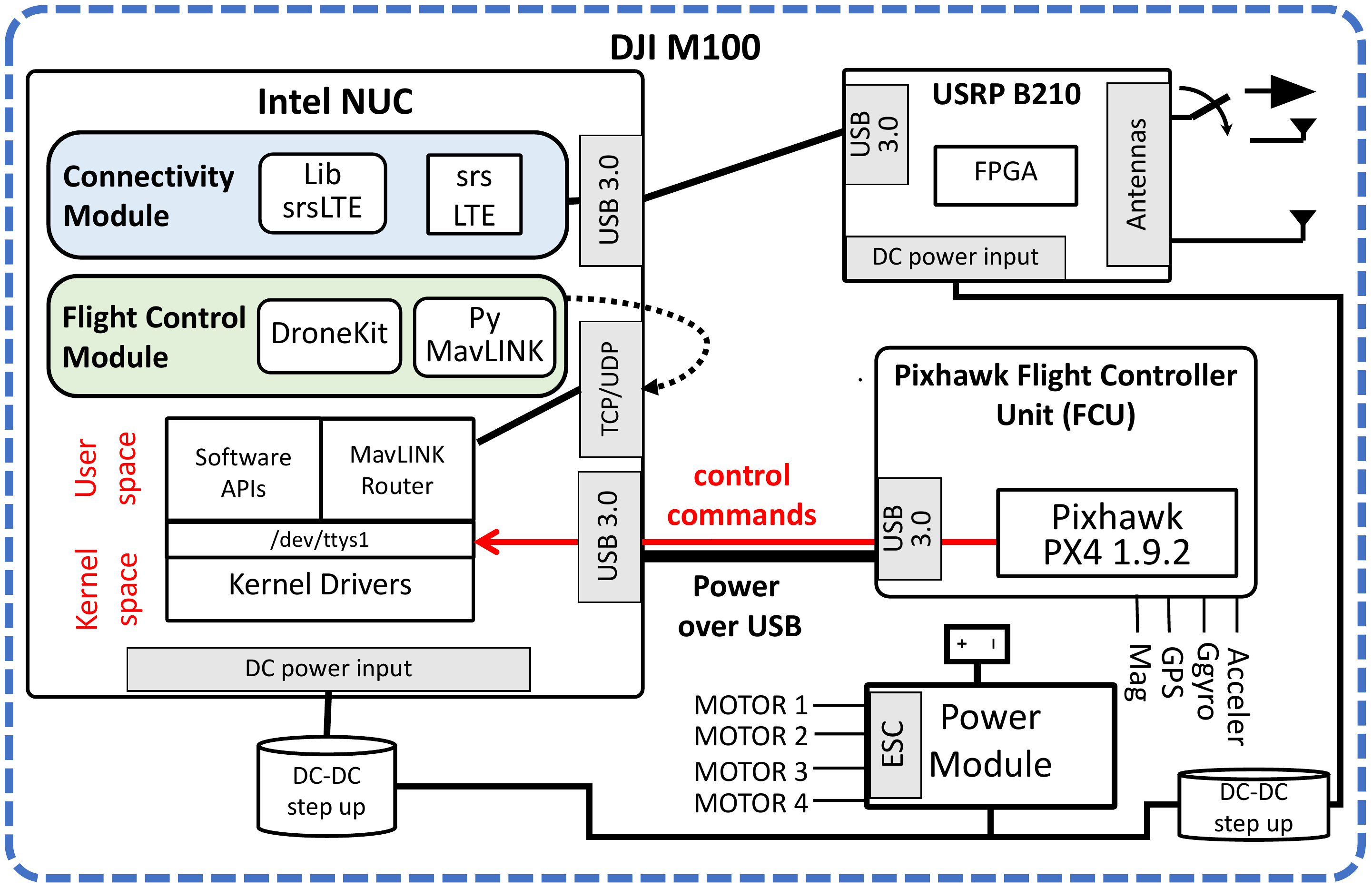}
  }
    \subcaptionbox{
  UAV-based UE model.  
  \label{fig:prototype_photo}
  }
  {
    \includegraphics[width=.55\columnwidth]
    {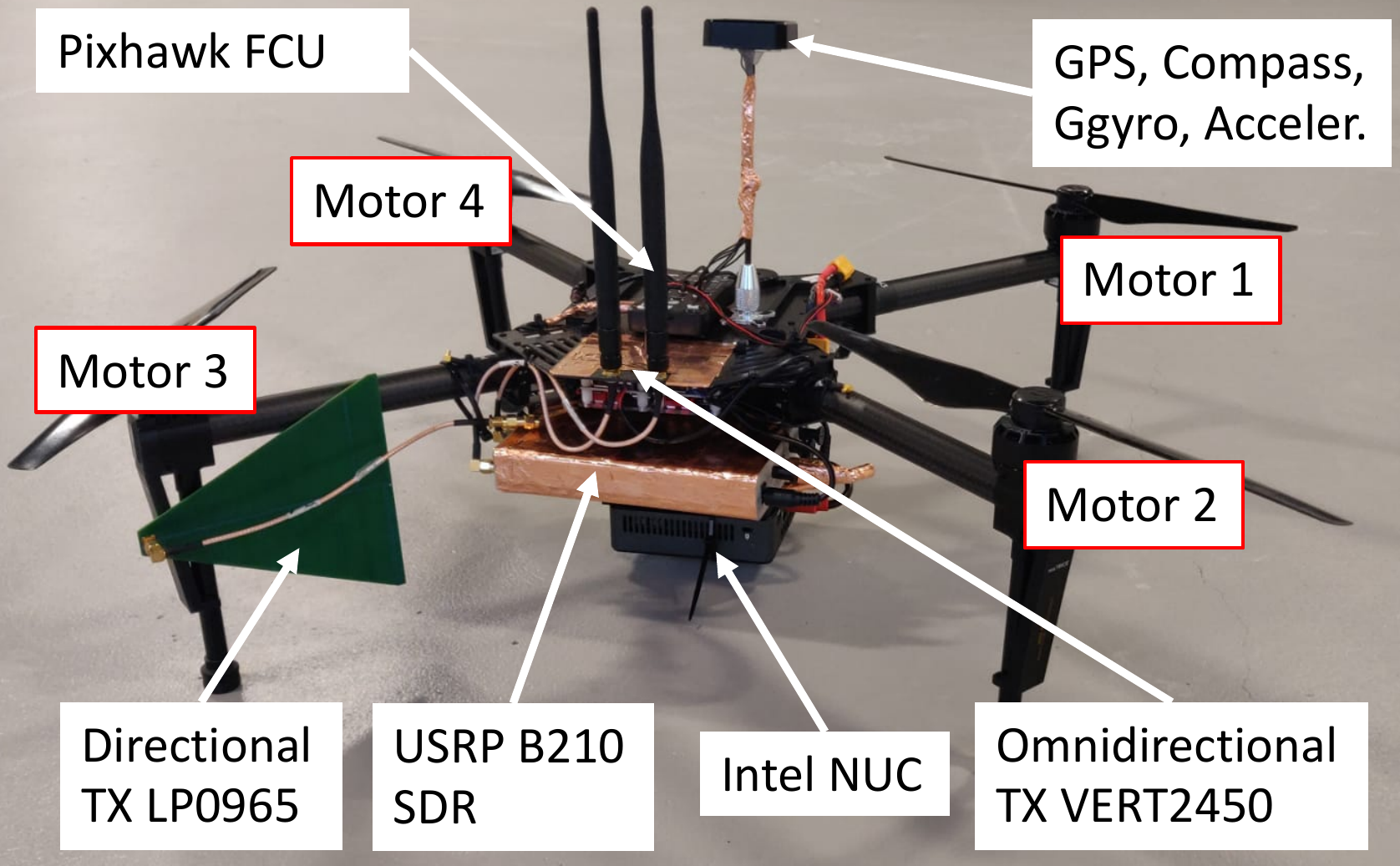}
  }
  \caption{UAV-based UE prototype}
  \label{fig:prototype}
\end{figure}

\section{Performance Evaluation}
\label{sec:evaluation}
In assessing the performance of the proposed control scheme, we start with an experimental evaluation on a full-stack, multi-cell LTE cellular network testbed.
Our LTE-based analysis adopts the same 5G low- and mid-band frequencies suggested for drone cellular communications (FR1) \cite{3gpprel15}.
To the best of our knowledge, this is the first real-world experiment of a connected-drone control system on a full-stack RAN testbed outdoors. 
Then, we extensively assess the performance of the proposed control scheme on a series of realistic large-scale deployments through simulation. In doing so, we utilize the commercial deployment topology and actual cell-load profile of a major US cellular carrier in several representative scenarios. 

\subsection{Small-scale Real-world Experiments}
\label{sec:experiments}
\textbf{\indent Experimental Setup.} For our experiments, we prototype a multi-eNB LTE cellular network testbed using srsLTE which is an open-source, full-stack implementation of LTE for \glspl{sdr}  \cite{gomez2016srslte, nikaein2014openairinterface}.
Our fully controllable LTE cellular testbed constitutes the following components. \newline 
\begin{itemize*} 
    \item \emph{Software stack:} The srsLTE suite comes with three modules, namely \texttt{srsepc}, \texttt{srsenb}, and \texttt{srsue}, that implement the  LTE-compliant protocol stacks for the core network, the eNB, and the UEs, respectively.
    \newline \item \emph{Host machines:} Each of the components above is hosted on an Intel NUC computer. 
    The Intel NUC is a commercial Mini PC, whose compact dimensions 
    and good computational capabilities (Intel Core i7 CPU with $32\:\mathrm{GB}$ RAM) make it particularly suitable even to be carried on board of an UAV.
    \newline \item \emph{Radio front-ends:} We use the \gls{usrp} B210, a high-bandwidth, high-dynamic range \gls{sdr} designed to operate from DC to $6\:\mathrm{GHz}$ frequency. 
\end{itemize*}

\begin{figure}[t!]
  \centering
  \subcaptionbox{
  Scenario 1: deployment.
  \label{fig:fly1deployment}
  }
  {
    \includegraphics[width=.7\columnwidth]
    {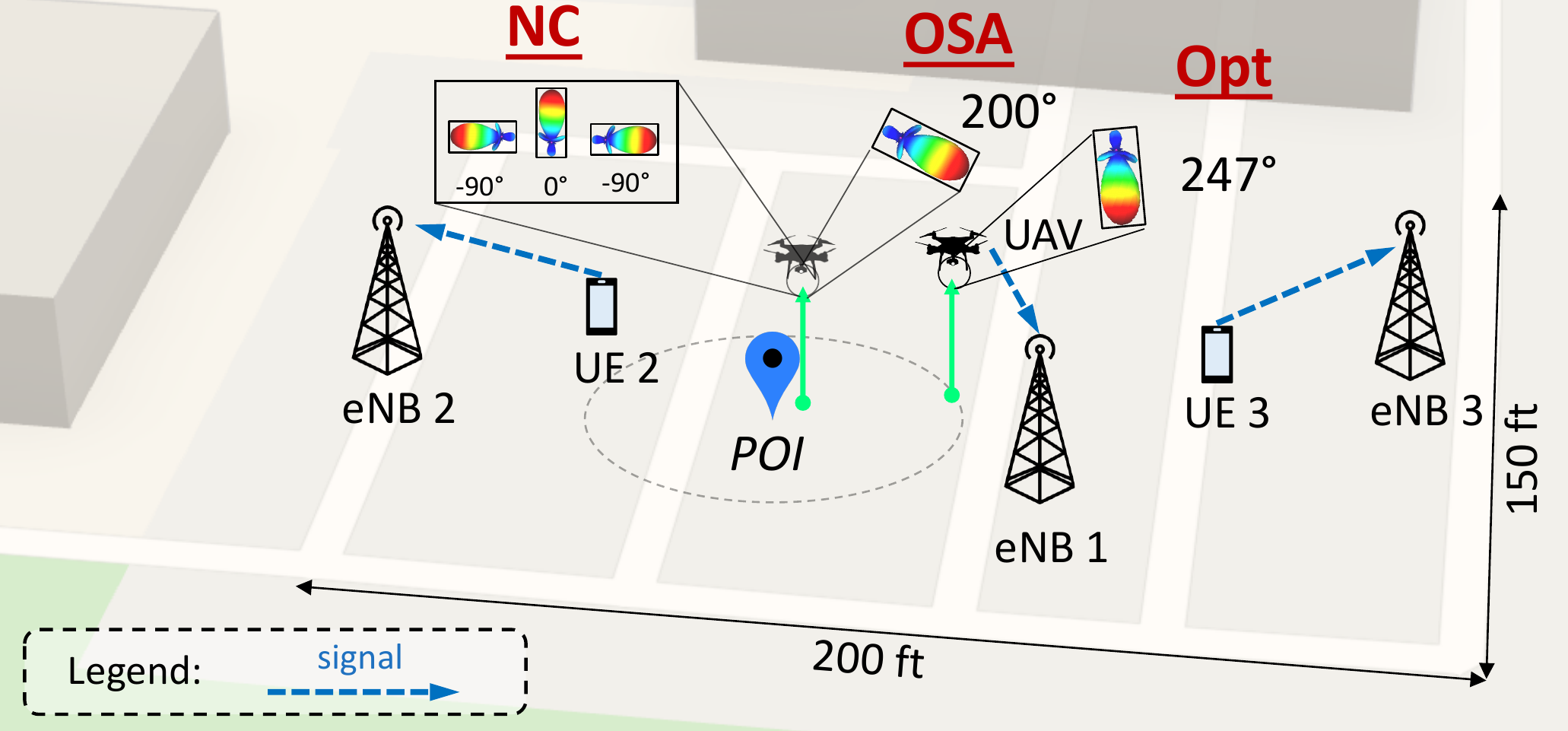}}
  \subcaptionbox{
    Scenario 1: Single-run performance. 
  \label{fig:fly1performance}
  }
  {
    \includegraphics[width=.6\columnwidth]
    {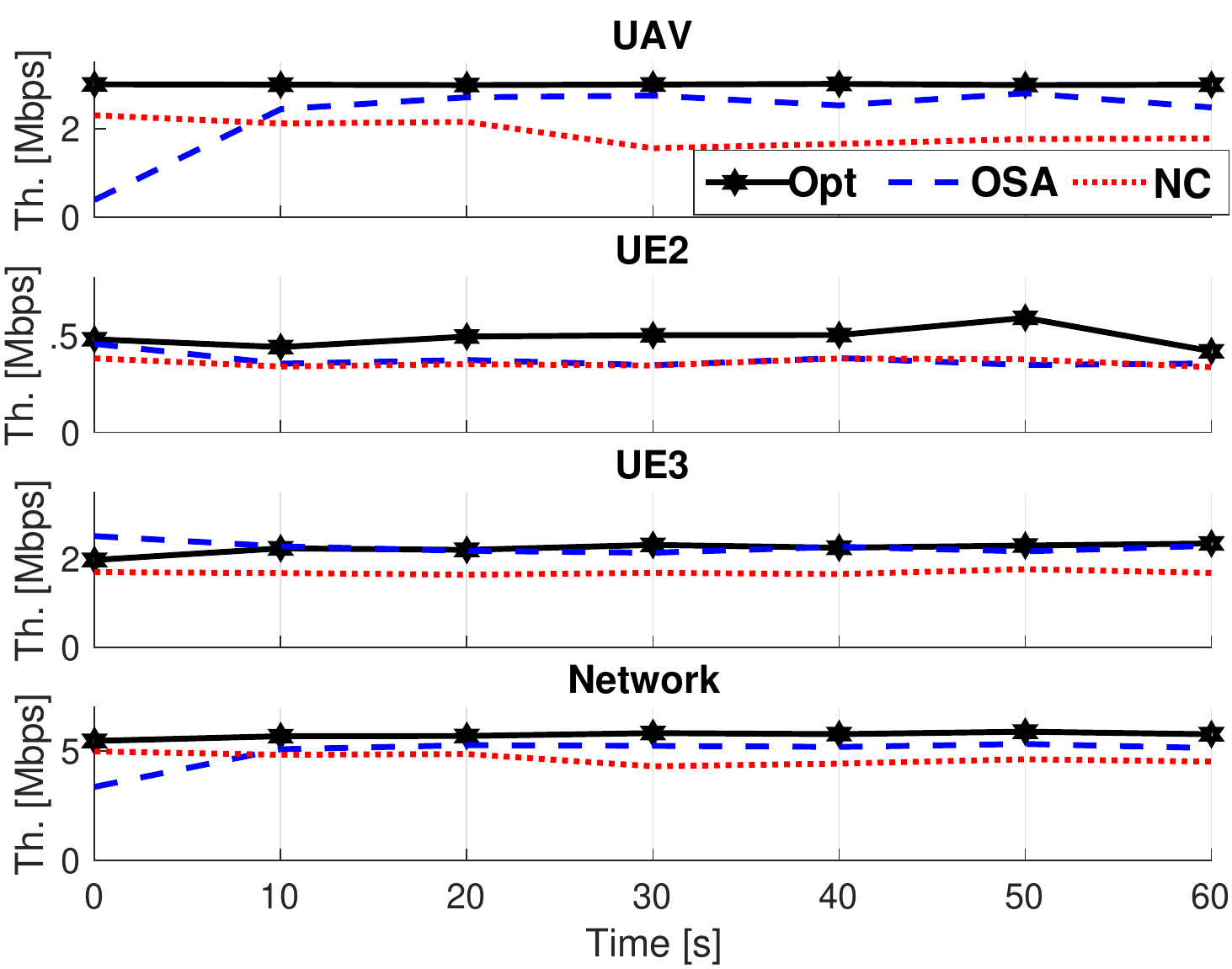}
  }
  \caption{Network testing scenario 1.}
  \label{fig:fly1}
\end{figure}

\begin{figure}[t!]
  \centering
  \subcaptionbox{
  Scenario 2: Deployment.
  \label{fig:fly2deployment}
  }
  {
    \includegraphics[width=.7\columnwidth]
    {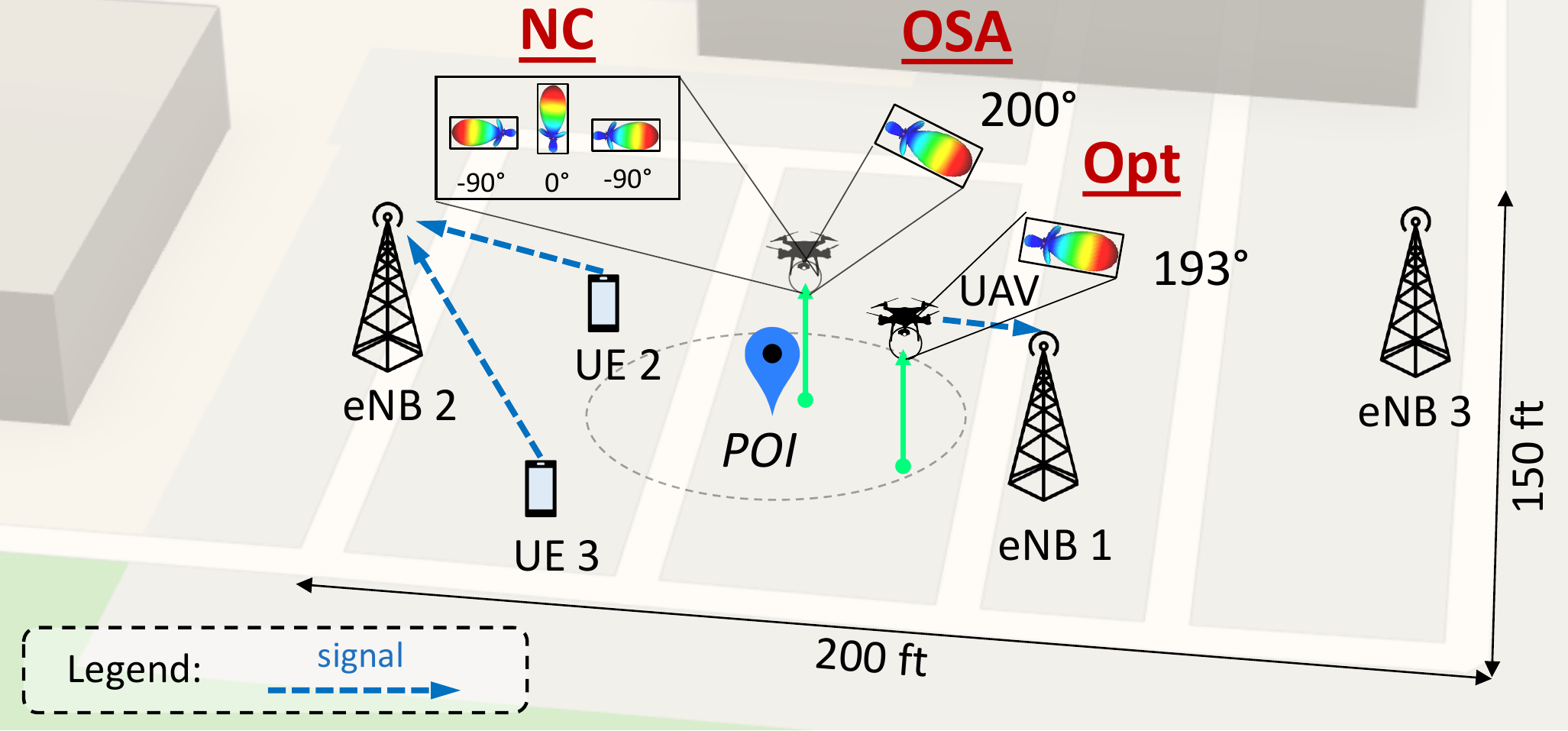}
  }
  \subcaptionbox{
    Scenario 2: Single-run performance.\vspace{0mm}
  \label{fig:fly2performance}
  }
  {
    \includegraphics[width=.6\columnwidth]
    {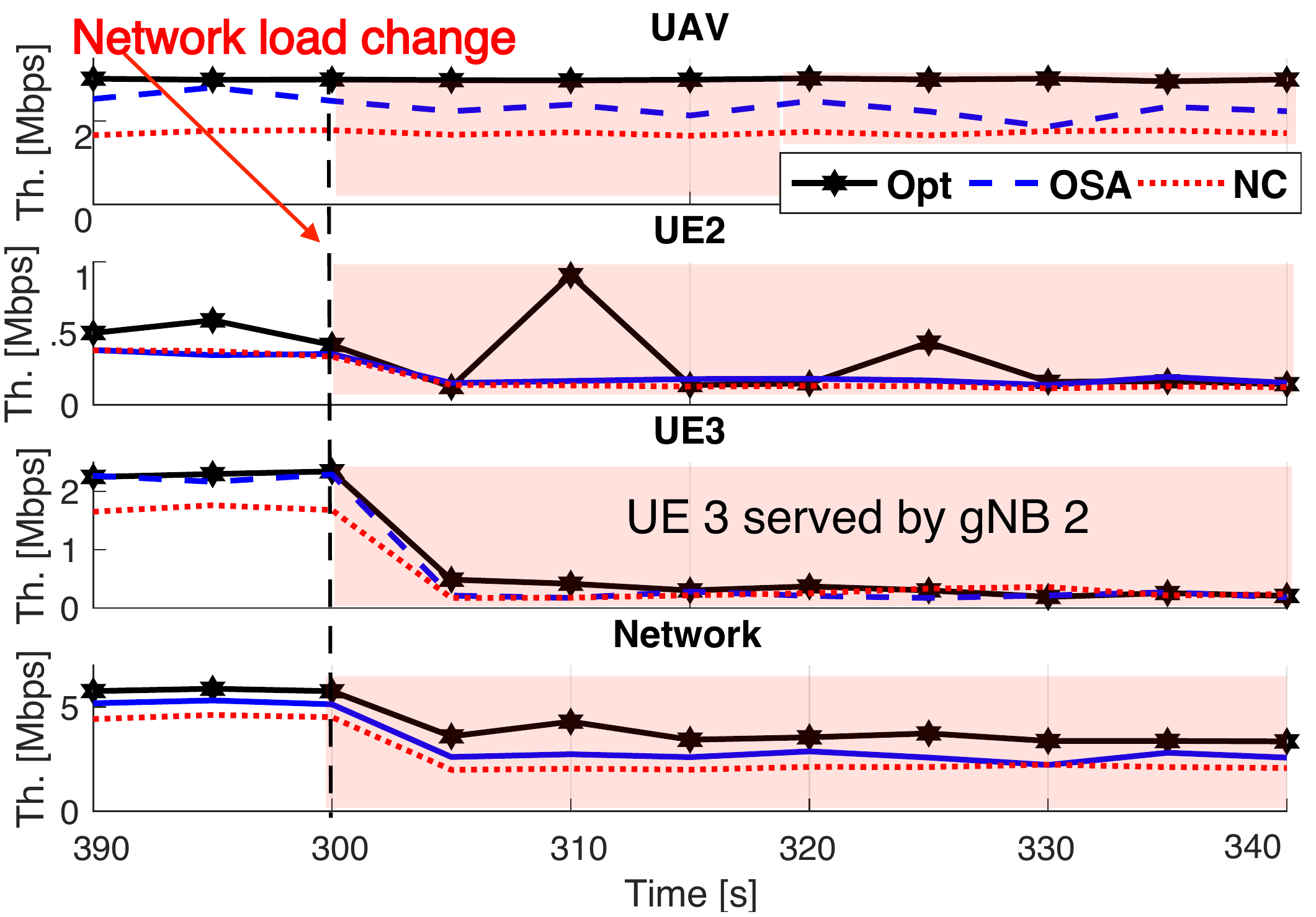}
  }
  \caption{Network testing scenario 2.}
  \label{fig:fly2}
  \vspace{0mm}
\end{figure}

The deployment topology of our network consists of 3 eNBs covering $37500\:\mathrm{ft}^2$ of outdoor space
and a mixed set of 3 UEs as illustrated in Figs. \ref{fig:motivation_topology}, \ref{fig:changeLocationDeployment},  \ref{fig:changeRotationDeployment},  \ref{fig:fly1deployment}, and  \ref{fig:fly2deployment}.
In our deployments, each eNB serves one UE that is located approximately $25\:\mathrm{ft}$ away, with the eNBs located around $150\:\mathrm{ft}$ apart.
The eNBs' antennas are placed at $10\:\mathrm{ft}$ from the ground using adjustable tripods, while the UE devices are placed at the ground level.
Each eNB and UE uses a pair of antennas VERT 2450 for transmission and reception over the air, whose radiation patterns can be found at \cite{vert2450}.
Further, UE 1 is tethered to a UAV as shown in Fig.~\ref{fig:prototype_photo}. In our proposed solution, UE 1 employs a directional log-periodic LP0965 antenna at the TX side as shown in Fig.~\ref{fig:prototype_photo}. The LP0965 module is a wide-angle log-periodic directional antenna with $90^{\circ}$ of horizontal aperture and $6\:\mathrm{dBi}$ of forward gain, whose specifications can be found at \cite{LP0965}.
We designed and assembled a UAV-based UE by mounting a companion computer (Intel NUC) and a USRP B210 \gls{sdr} on a DJI M100 drone.  
For the drone flight and motion control, we employed the Pixhawk flight controller board and the PX4 flight control firmware. 
The schematics of the prototyped aerial UE are illustrated in Fig.~\ref{fig:schematics}.
Lastly, to mimic the blockage and shadowing effect of typical cellular deployments in our experimental motivation section (see Sec. \ref{sec:motivation}, Fig. \ref{fig:motivation_topology}) we use a large van positioned right behind the UAV. This way, when the UAV is on the ground, it will experience propagation patterns similar to traditional ground UEs, that is, being in full LoS with at most one eNB only.

The 3-eNB network operates on LTE Band $7$ where the uplink and downlink central frequencies are at $2535\mhz$ and $2655\mhz$, respectively (which also corresponds to 5G NR band $7$). 
To achieve a more robust testing deployment, we intentionally limit the operational bandwidth of the uplink and downlink directions to $3\:\mathrm{MHz}$ ($15$ \gls{prb}s). This reduces the pressure on SDR's host-based baseband processing, prevents user disassociation, and limits the interference to and from commercial cellular networks that operate on the same band.
We reduce the complexity of our system by connecting a single UE to each of the eNBs. For our performance measurements, we saturate the uplink channel at each UEs so that each UE uses up all the uplink PRBs from its serving eNB. 
To do so, we implement an \texttt{iperf} server at each eNB and an \texttt{iperf} client at each UE. 
Then, we select UDP over TCP as the transport protocol for its lack of congestion control, which might interfere with our performance evaluation. 
This configuration is designed to mimic the uplink inter-cell interference conditions of a commercial cellular network in which multiple UEs collectively occupy the whole spectrum band of their serving cells and create uplink inter-cell interference.
We implement the RAN controller on a laptop connected to the network. Given the limited CPU core pool and unexceptional parallelization features of our machine, the mathematical solver takes around $25$ seconds to find a solution to the CDCP. Accordingly, we adopt an optimization interval $\mathcal{T} = 5$ minutes for our real-world experiments in a controlled environment. On commercial hardware with a larger core pool and higher degrees of parallelization, the solver run-time is expected to be in the order of milliseconds, which can be safely assumed negligible  when $\mathcal{T}$ is in the order of minutes.
In the following, we experimentally evaluate the effectiveness and adaptability of our proposed connected-drone control solution.
Last, for safety reasons, we kept the UAV’s altitude fixed to $10\:\mathrm{ft}$ in our real-world experiments (while we do optimize this parameter in the large-scale simulations presented later).
Finally, in both the small-scale experiments and the large-scale simulations presented, we implement our control system using the python's {\tt scipy.optimize} suite and calculate the control solution using the {\tt scipy.minimize.basinhopping} method \cite{basinhopping}.

\begin{figure}[b!]
\centering
\includegraphics[width=0.8\columnwidth]{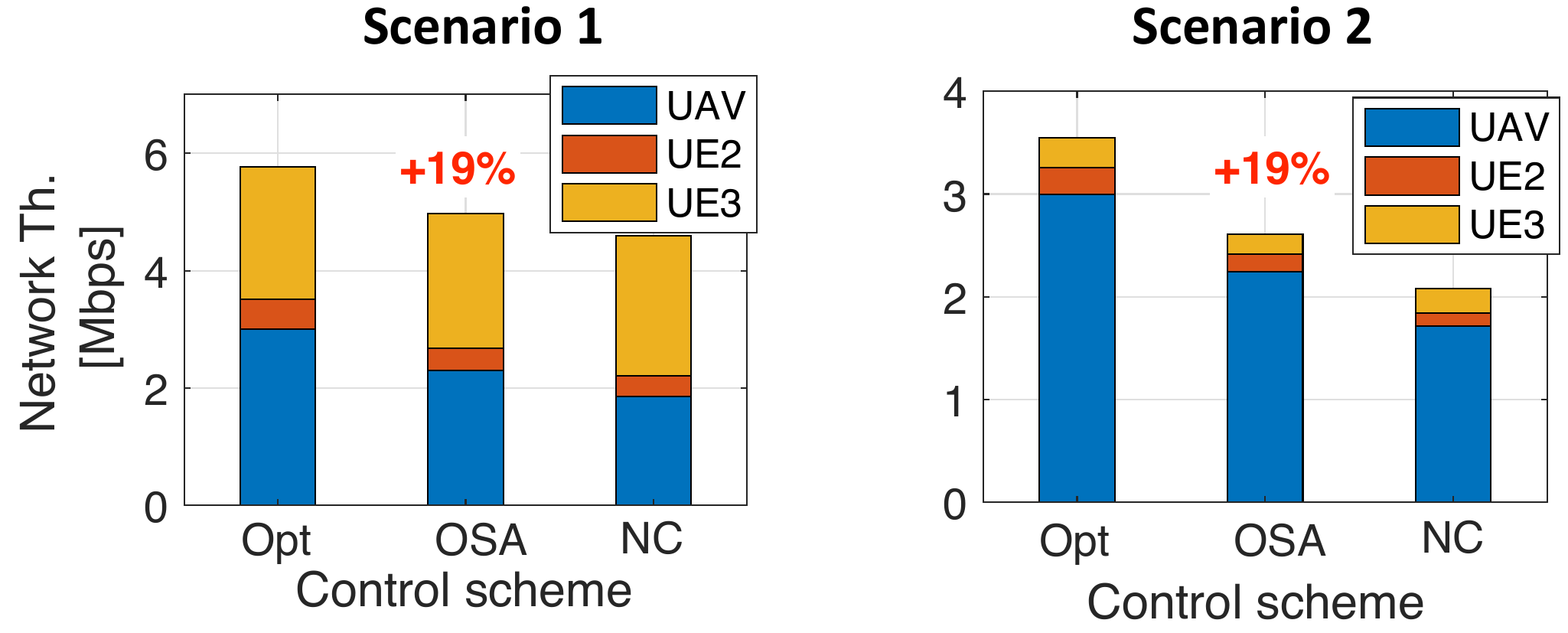}
\caption{Average experimental performance results.}
\label{fig:averageexperimental}
\end{figure}

\textbf{Effectiveness.} 
In testing the proposed control solution, we start from the deployment topology reported in Fig. \ref{fig:fly1}. This consists of 3 eNBs --- eNB 1, eNB 2, and eNB 3; and 3 UEs --- the UAV, UE 2, and UE 3. 
Due to the dominant LoS conditions of our testing scenario, the UEs serving cell is selected as the nearest one for all our real-world experiments.
\textit{Herein, we assess the performance of the proposed Open-RAN-based control (\ul{Opt}) against two control schemes that do not access comprehensive RAN infrastructure information and operate traditional intra-cell performance optimization.} 
These other two control schemes are BS-constrained, meaning that they aim at maximizing the uplink throughput performance of each BS, independently. These are:
\begin{enumerate*}[label=(\roman*)]
\item \ul{\textit{No Control} (NC)}, which adopts the POI as optimal drone location and a fixed, random transmission directionality. For this, we average the measurement across the $180^{\circ}$ in the direction of the serving BS; and
\item \ul{\textit{Optimal Serving Angle} (OSA)}, which envisions the \gls{poi} as optimal UAV's location and the direction toward the serving BS as TX directionality.
\end{enumerate*} 
For each control scheme, we measure the individual users' and the overall network performance.

In our experiment, the drone specifies a POI in between cell 1 and cell 2 demanding a maximum distance of $30\:\mathrm{ft}$ from it. For these scenarios, we specify a minimum drone's QoS of $2\mbps$. 
The deployment scenario for this experiment is illustrated in Fig \ref{fig:fly1deployment}. 
We run a 1-minute-long experiment to assess the proposed control scheme and the two baselines, and report the individual UEs throughput together with the aggregate network performance in Fig \ref{fig:fly1performance}.
When adopting the Open-RAN-based control solution, the drone achieves higher throughput and overall better network performance as reported in Fig.~\ref{fig:fly1performance}.
Over 10 1-minute long experiments, the UAV, UE 2, and UE 3 achieve $+0.69\mbps$, $+0.14\mbps$, and $-0.04\mbps$ with respect to OSA, and $+1.14\mbps$, $+0.16\mbps$, and $-0.13\mbps$ with respect to NC.
The average performance results are reported in Fig.~\ref{fig:averageexperimental} (left).
Overall, our approach improves the network performance by $+16\%$ and $+25\%$ with respect to OSA and NC while guaranteeing the UAV's uplink QoS and application requirement distance from the POI. 

\textbf{Adaptability.} Under the Open-RAN-based control solution, the drone operates with the guarantee of satisfactory QoS and optimized network performance. This holds for the whole time interval $\mathcal{T}$ as long as the UAV's application requirements stay unchanged (POI and QoS).
At every new time interval $\mathcal{T}$, the control system retrieves new network parameters from the Open-RAN architecture and proceeds to the calculation of a new optimal solution.
We evaluate the adaptability of the proposed control system to changing network conditions by inducing a network load change in our deployments.
To induce a significant change in cell load, we move UE 3 from cell 3 to cell 2. This way, we double the cell load on eNB 2 and reduce the load on eNB 3 to zero, which induces a major load shift in our system.
This perturbation emulates stationary geographical traffic shifts throughout the day, such as from business to residential districts after office hours.
With this experiment, we demonstrate how the proposed control approach allows the computation of a new operational point at every interval $\mathcal{T}$ by dynamically adapting to the changing network conditions. Accordingly, we assess the performance of our solution, as well as the two BS-constrained optimization schemes NC and OSA after a network change. 
Unlike the proposed Open-RAN control solution, these control schemes are unable to implement a network-wide optimal policy, and can only focus on intra-cell performance optimization. This, in turn, disregarding the interference implications on the rest of the network and results in poorer network performances.
Fig.~\ref{fig:fly2} illustrates the new network deployment (Scenario 2) and the corresponding network performance analysis. By dynamically adapting to the new network conditions, the proposed control scheme calculates and implements a new optimal solution that guarantees superior performance for both the UAV and the other ground UEs as reported in Fig. \ref{fig:fly2performance}. 
In this new scenario, the UAV, UE 2, and UE 3 achieve $+0.53\mbps$, $+0.1\mbps$, and $+0.1\mbps$ with respect to OSA, and $+1.28\mbps$, $+0.14$, and $+0.06\mbps$ with respect to NC. Overall, the proposed Open-RAN control solution achieves $+36\%$ and $+52\%$ network throughput over the OSA and NC control schemes.

\begin{figure}[t!]
\centering
\includegraphics[width=0.5\columnwidth]{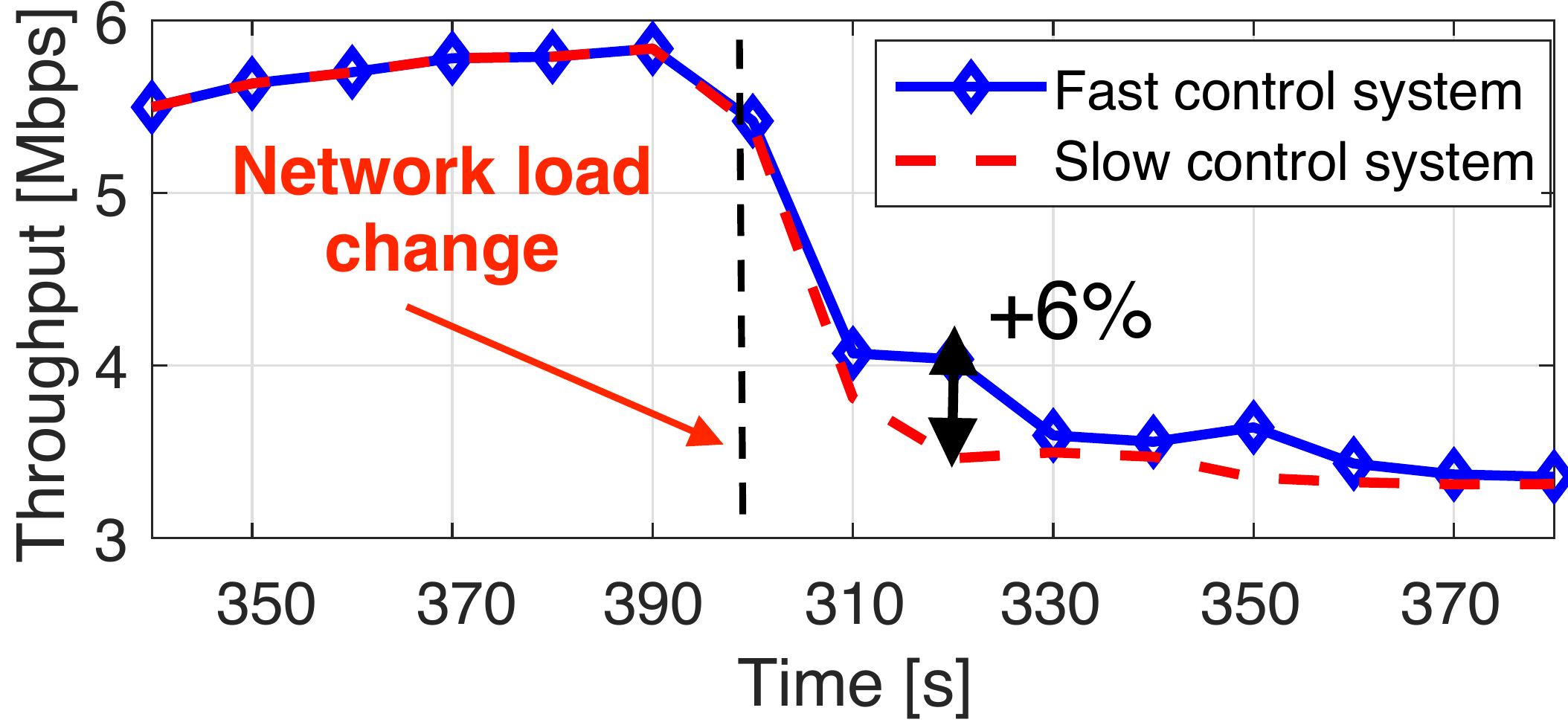}\vspace{0mm}
\caption{Performance toll for slowly-reacting control systems.}
\label{fig:dynamicsol}
\vspace{0mm}
\end{figure}

\textbf{Flexibility:}
Last, we are interested in quantifying the benefits of adopting a proactive network control system that dynamically adapts to the changing network conditions in neighboring cells at every interval $\mathcal{T}$.
As mentioned above, the proposed control system takes full advantage of Open-RAN architectures to retrieve information from neighboring cells every time interval $\mathcal{T}$, and implement network-wide optimal solutions that satisfy the uplink QoS of the UAV while optimizing the performance of the ground UEs attached to the RAN.
This new system is thus different from traditional optimization policies that are constrained to individual BSs (the control scheme OSA, and OLA and OSLA that we will introduce later).
Additionally, our system allows solving the formulated CDCP with a time granularity $\mathcal{T}$ that can be dimensioned accordingly to the dynamics of the network deployment, such as users' mobility patterns.
Our control solution allows different systems to implement different optimization cycle granularity, depending on the specific needs. For example, slowly changing network environments might allow longer $\mathcal{T}$, while highly dynamic network deployments with many substantial traffic load changes per day should adopt faster control loop intervals.
Here, we quantify and report the benefits of adopting a faster optimization interval $\mathcal{T}$ with respect to a slower optimization interval.
We report the performance toll paid by a slower optimization system with respect to a faster optimization system by quantifying the network performance drop that occurs after a major traffic load change.
To support the need for dimensioning $\mathcal{T}$ according to the network system's dynamics, we compare the performance of two control systems right after a major traffic load change. As an example, we compare a system with shorter optimization interval ($\mathcal{T} = 5$ minutes) with respect to a system with a longer optimization interval ($\mathcal{T} = 10$ minutes).
Fig.~\ref{fig:dynamicsol} reports a network throughput drop of $6\%$ for the system with a longer optimization interval. This performance gap translates into a $120\mbps$ uplink throughput loss per second on a commercial network every time there is a major traffic load change.
As such traffic load changes might occur several times a day in highly dynamic network deployments, it is more appropriate for these environments to select a shorter value of $\mathcal{T}$ and have a more reactive network control system.

\begin{figure*}[t!]
\centering
\includegraphics[width=1\textwidth]{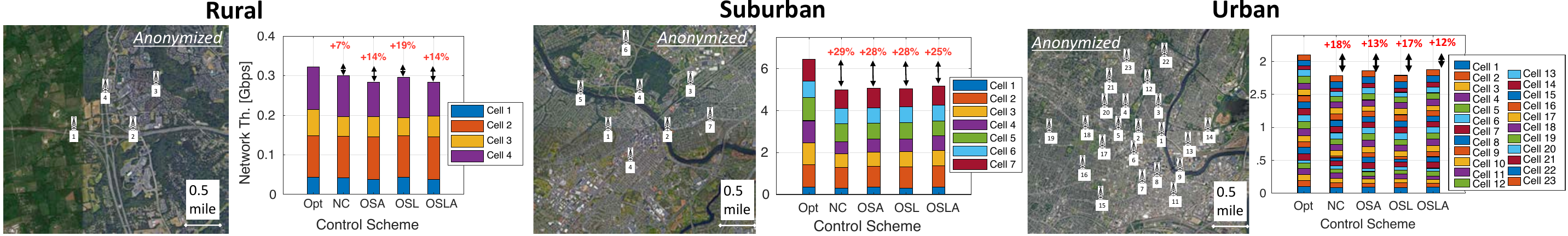}\vspace{0mm}
\caption{\ul{\textit{Anonymized}}\textsuperscript{1}deployment topologies and performance analysis for the three large-scale scenarios.} \vspace{0mm}
 \label{fig:largescale}
\end{figure*}

\vspace{0mm}
\subsection{Large-scale Simulation Analysis}
\label{sec:simulaton}

\textbf{\indent Scalability.}
Here, we assess the performance of our solution on a series of large-scale, realistic cellular deployment scenarios. 
For this task, we simulate three commercial deployment topologies of a major US cellular carrier: rural, suburban, and urban. 
These differ in terms of scale, inter-cell distances, traffic loads, and cell deployment density. 
Accordingly, we assume that $\mathcal{T}$ is correctly dimensioned to respect the specific network deployment's dynamics, and we make the wireless carrier's commercial cell-load profiles available to our solver. 
In line with the CDCP discussion, we assume our simulated control system to be oblivious with respect to the individual ground users' traffic rates (i.e., the $r_i$ terms in CDCP (3)--(6)) and thus treat the cell-load profiles in terms of the average number of RRC connected UEs (i.e., the $\widetilde{w}_j$ terms in CDCP (6)--(10)). Accordingly, in the following simulation results, we report the overall network performance in terms of overall uplink network capacity calculated by our control solution. 
This is an upper bound to the overall network throughput performance that ultimately depends on the individual user's traffic requirements, which are non-observable by the CDCP solver.

An anonymized illustration of the three deployment scenarios is reported in the top part of Fig.~\ref{fig:largescale}.
This assessment adds a high degree of realism to our evaluation and is aimed at quantifying the gains of the proposed control approach on a real production network.

For our large-scale evaluation, we compare the performance of the Open-RAN-based control scheme (\ul{Opt}) with \ul{NC}, \ul{OSA}, and two additional BS-constrained control schemes: i) \ul{\textit{Optimal Serving Location} (OSL)}, which favors the drone's application data-rate by envisioning the drone hovering as close as possible to the serving BS with a fixed random transmission directionality (we average the results over the $180^{\circ}$ in the direction of the serving BS); and ii) \ul{\textit{Optimal Serving Location and Angle} (OSLA)}, which maximizes the drone's application data-rate envisioning the drone hovering as close as possible to the serving BS with transmission direction pointing at the serving BS.
While the proposed Open-RAN-based solution (Opt) can employ network-wide deployment and cell-load information, the 4 BS-constrained control schemes (NC, OSA, OSL, OSLA) perform traditional intra-cell optimization logic, and are thus unable to capture and address the impact of the UAV on terrestrial users in the surrounding cells.
For all the deployment scenarios, we assess the performance of the above control approaches for $20$ different random POIs.
In solving the CDCP for these realistic scenarios, we consider as  drone's serving cell $j_u$ the cell with one with the strongest signal path according to published aerial channel models \cite{amorim2017radio}. 
Without loss of generality, we assume all cells operating on the same carrier frequency and use commercial cell load profiles for a given band.
The \ul{\textit{anonymized}}\footnote{\label{ft:nda}Due to our NDA with the cellular carrier, we cannot report the actual cell load profile or cell deployment. For the sake of completeness, we however report similar cell load profiles and deployments per density and scale.} cell load profile for the three deployments is the following. 
Rural: $57$, $108$, $36$, $147$ UEs; 
Suburban: $48$, $135$, $27$, $36$, $357$, $261$, $168$ UEs; 
Urban: $120$, $30$, $21$, $57$, $129$, $195$, $60$, $27$, $54$, $60$, $126$, $99$, $75$, $45$, $30$, $123$, $90$, $18$, $90$, $86$, $33$, $198$, $78$ UEs. 
Lastly, we express the UAV's uplink requested QoS as $5\mbps$ (\eg to support a 1080p and 60 fps stream), a maximum tolerable distance from POI of $500\:\mathrm{ft}$, and a minimum SINR at the neighboring cells of $25\db$. We constrain the minimum drone's flight altitude according to the deployment's average building height ($65\ft$ for rural and suburban, $165\ft$ for urban).

For each experiment, we measure the per-cell aggregate ground users' uplink throughput and the overall compound uplink network capacity (in $\mathrm{Gbps}$). We report the average performance for the three scenarios and the 5 control schemes in Fig. \ref{fig:largescale}.
The proposed Open-RAN-based control solution outperforms all the BS-constrained control schemes in all the considered deployment scenarios. 
The proposed approach reports uplink network capacity improvements over the second-best performer as high as $23\mbps$, $128\mbps$, and $223\mbps$ for rural, suburban, and urban deployments. These translate into an overall increased network capacity of $7\%$, $25\%$, and $12\%$. 
On average, the proposed control system based on Open-RAN architectures achieves $+19\%$ uplink network capacity increase over the other control schemes that implement traditional intra-cell optimization approaches on real, commercial network deployments, while satisfying target location and HQ video-streaming requirements to the UAV.

\section{Related Work}
\label{sec:relatedwork}
While interference mitigation for exposed terminals in wireless networks is a well-studied subject \cite{kumar2006medium, vilzmann2006survey, bonati2020cellos}, cellular-connected drones present some fundamental differences. 
A cellular-connected drone hovers at unprecedented and unprotected altitudes for a UE (different from pedestrians or users in tall buildings), it is highly mobile (unlike users in hot air balloons or high up a hill), and its applications demand high data-rates (unlike high-altitude sensors). 
Thus, previous interference cancellation studies are hardly applicable to this scenario, which remains under-explored.
The difficulty of extending the ground-tailored wireless infrastructure to UAVs and the resulting A2G and ground-to-air (G2A) interference conditions have been highlighted in several works \cite{azari2017coexistence, zeng2018cellular, ivancic2019flying, stanczak2018mobility,  mei2019cellular, lin2018sky, nguyen2018ensure, van2016lte}.
Some produced propagation aerial models, setting the ground for future aerial network research \cite{khuwaja2018survey, polese2020experimental}. 
Some other works proposed interference mitigation solutions for cellular-connected UAVs. 
\cite{izydorczyk2018experimental} proposed a multi-antenna interference cancellation for the downlink channel while \cite{mei2019uplink} proposed a cooperative non-orthogonal multiple access (NOMA) technique to mitigate the uplink interference at ground BSs.
These approaches fail to provide a proactive solution for interference-prone A2G communications, leaving the burden of interference cancellation to complex reactive BS cooperation schemes.  
The use of directional transmitters at the UAVs is suggested in \cite{amorim2018measured, lyu2017blocking} to improve both uplink and downlink communications. 
Directional transmission for UAV-based UEs is also proposed in \cite{BertizzoloHotmobile20}, where Bertizzolo et al. design a cooperative control mechanism to optimally control the trajectory and the orientation of the drone's transmission to mitigate the interference toward neighboring cells when the drone relocates from one way-point to another.
These approaches focus on characterizing the benefits of directional transmitters and motion control without considering the specific bandwidth and application-specific requirements.
On the other hand, other works have explored UAV location control for wireless network optimization outside the cellular paradigm \cite{cheng2021learning, BertizzoloInfocom20SwarmControl, sundaresan2018skylite, chakraborty2018skyran, sheshadri2020skyhaul}. 
Unlike these works, we contextualize UAV-based UE broadband connectivity problem into the Open-RAN paradigm and approach it from an ISP's standpoint. 
Here, we exploit the architectural advantage of Open-RAN architectures to address the need for a drone-specific control solution that supports location and data-rate requirements for aerial video streaming applications.

\section{Conclusions and Final Remarks}
\label{sec:conclusion}
We introduced a new closed-loop control solution for Open RANs to support drone-based video streaming applications on commercial 5G cellular networks. 
In the proposed control system, a UAV-based UE contacts a service at the Open-RAN controller for a satisfactory location and transmission directionality that support the application-specific location and QoS requirements. Exploiting the architectural advantage of Open-RAN controllers, the service computes and returns a solution that maximizes the overall uplink network performance and matches the UAV's required QoS. 
The proposed approach features \textit{5G compatibility}, \textit{low overhead}, and \textit{user privacy}. We assess the proposed solution on a real-world, outdoor, multi-cell RAN testbed. 
Further, we perform at-scale simulations employing the commercial cell deployment topologies and cell load profiles of a major US carrier. We prove \textit{effectiveness} of the proposed control system (average $19\%$ network capacity gain), \textit{adaptability} to dynamic network conditions, and \textit{scalability} to real-world deployments.
We conclude this paper with a few final remarks.
\newline \textbf{\textit{Standard APIs.}}
In this article, we proposed a control system that enables drone-sourced video streaming without requiring any modification to the 5G protocol stack. 
Full integration of the proposed solution into commercial applications that guarantees the properties of low-overhead and user-privacy asks for standardized APIs. This will enable vendor-agnostic support and help the development of drone-enabled cellular networks.
Similarly, Open-RAN compliant signaling procedures (\eg based on O-RAN) are needed for cell-load information retrieval.
\newline \textbf{\textit{Drone Mobility.}}
To preserve users' privacy, the proposed solution does not implement mobility control or tracking. The implementation of the computed new location solution is instead left to the drone. This makes our approach, if you will, easily extendable and compatible with other trajectory control solutions and  previous work on the subject, \eg \cite{BertizzoloHotmobile20}.
\newline \textbf{\textit{Directional Transmitters.}}
Last, we care to remark that the proposed approach is directional transmitter-agnostic. While wide-angle directional TXs today are a more scalable and cost-effective solution for UAVs in FR1, the proposed approach will easily support highly-directional communications in FR2 and represents a ready-to-go solution for when NextG networks will support high directionality in FR1. 
\newline \textbf{\textit{Future Work.}}
In the future, we will take these research directions:
\newline\textit{Multiple UAVs}: We intend to extend the current solution support to multiple UAVs. For this, we will model the aerial UAV-to-UAV interference and, accordingly, extend the current CDCP formulation to support multi-drone optimization functions. 
\newline\textit{Specific Application Requirements}: We will also focus on how to satisfy video streaming applications with specific requirements such as particular shooting angles and optimize the scenarios where surrounding objects or buildings could obstruct the view of the POI from certain locations in space.

\section{Acknowledgments}
This work was partially supported by the U.S. National Science Foundation under Grant CNS-1923789, the U.S. Office of Naval Research under Grant N00014-20-1-2132, and an AT\&T Virtual University Research Initiative (VURI).


\bibliography{biblioTMC.bib}

\begin{thebibliography}{10}

\bibitem{3gpprel15}
{3GPP TR 36.777, Enhanced LTE support for aerial vehicles}, 2019.
\newblock {ftp://www.3gpp.org/specs/archive/36\_series/36.777}.

\bibitem{oran}
O-ran ric specifications.
\newblock {https://www.o-ran.org/specifications}, 2019.

\bibitem{vodafone}
Vodafone and ericsson trial establishes automated flight paths for connected
  drones, 2020.
\newblock
  {https://www.vodafone.com/news-and-media/vodafone-group-releases/news/vodafone-ericsson-safe-passage-drones}.

\bibitem{amazon}
Amazon.
\newblock Amazon prime air, 2016.

\bibitem{amorim2017radio}
Rafhael Amorim, Huan Nguyen, Preben Mogensen, Istv{\'a}n~Z Kov{\'a}cs, Jeroen
  Wigard, and Troels~B S{\o}rensen.
\newblock Radio channel modeling for uav communication over cellular networks.
\newblock {\em IEEE Wireless Commun. Lett.}, 6(4):514--517, 2017.

\bibitem{amorim2018measured}
Rafhael Amorim, Huan Nguyen, Jeroen Wigard, Istv{\'a}n~Z Kov{\'a}cs, Troels~B
  S{\o}rensen, David~Z Biro, Mads S{\o}rensen, and Preben Mogensen.
\newblock Measured uplink interference caused by aerial vehicles in lte
  cellular networks.
\newblock {\em IEEE Wireless Communications Letters}, 7(6):958--961, 2018.

\bibitem{azari2017coexistence}
Mohammad~Mahdi Azari, Fernando Rosas, Alessandro Chiumento, and Sofie Pollin.
\newblock Coexistence of terrestrial and aerial users in cellular networks.
\newblock In {\em 2017 IEEE Globecom Workshops (GC Wkshps)}, 2017.

\bibitem{BertizzoloInfocom20SwarmControl}
Lorenzo Bertizzolo, Salvatore. D'oro, Ferranti Ludovico, Leonardo Bonati,
  Emrecan Demirors, Zhangyu Guan, Tommaso Melodia, and Scott Pudlewski.
\newblock {SwarmControl: An Automated Distributed Control Framework for
  Self-Optimizing Drone Networks}.
\newblock In {\em Proc. of IEEE Conf. on Computer Communications (INFOCOM)},
  pages 1768--1777, June 2020.

\bibitem{BertizzoloHotmobile20}
Lorenzo Bertizzolo, Tuyen~X. Tran, Brian Amento, Bharath Balasubramanian,
  Rittwik Jana, Hal Purdy, Yu~Zhou, and Tommaso Melodia.
\newblock Live and let live: Flying uavs without affecting terrestrial ues.
\newblock In {\em Proc. of the 21st Intl. Workshop on Mobile Computing Systems
  and Applications (HotMobile)}, Austin, TX, USA, March 2020.

\bibitem{bonati2020cellos}
Leonardo Bonati, Salvatore D'Oro, Lorenzo Bertizzolo, Emrecan Demirors, Zhangyu
  Guan, Stefano Basagni, and Tommaso Melodia.
\newblock {CellOS}: Zero-touch softwarized open cellular networks.
\newblock {\em Computer Networks}, 180, June 2020.

\bibitem{bonati2020open}
Leonardo Bonati, Michele Polese, Salvatore D'Oro, Stefano Basagni, and Tommaso
  Melodia.
\newblock Open, programmable, and virtualized 5g networks: State-of-the-art and
  the road ahead.
\newblock {\em Computer Networks}, 2020.

\bibitem{boubrima2020robust}
Ahmed Boubrima and Edward~W Knightly.
\newblock Robust mission planning of uav networks for environmental sensing.
\newblock In {\em ACM Workshop on Micro Aerial Vehicle Networks, Systems, and
  Applications (DroNet)}, pages 2--1, 2020.

\bibitem{buczek2021wireless}
John Buczek, Lorenzo Bertizzolo, Stefano Basagni, and Tommaso Melodia.
\newblock {What is A Wireless UAV? A Design Blueprint for 6G Flying Wireless
  Nodes}.
\newblock {\em arXiv preprint arXiv:2110.02472}, 2021.

\bibitem{chakraborty2018skyran}
Ayon Chakraborty, Eugene Chai, Karthikeyan Sundaresan, Amir Khojastepour, and
  Sampath Rangarajan.
\newblock Skyran: A self-organizing lte ran in the sky.
\newblock In {\em Proc. of the 14th Intl. Conf. on emerging Networking
  EXperiments and Technologies}, pages 280--292, 2018.

\bibitem{cheng2021learning}
Hai Cheng, Lorenzo Bertizzolo, Salvatore D’oro, John Buczek, Tommaso Melodia,
  and Elizabeth~Serena Bentley.
\newblock {Learning to fly: A distributed deep reinforcement learning framework
  for software-defined uav network control}.
\newblock {\em IEEE Open Journal of the Communications Society}, 2:1486--1504,
  2021.

\bibitem{cnn}
CNN.
\newblock Cnn cleared to test drones for reporting.
\newblock Technical report, 2015.
\newblock {https://money.cnn.com/2015/01/12/technology/cnn-drone/}.

\bibitem{gavrilovska2020cloud}
Liljana Gavrilovska, Valentin Rakovic, and Daniel Denkovski.
\newblock From cloud ran to open ran.
\newblock {\em Wireless Personal Communications}, pages 1--17, 2020.

\bibitem{gomez2016srslte}
I.~Gomez-Miguelez, A.~Garcia-Saavedra, P.D. Sutton, P.~Serrano, C.~Cano, and
  D.J. Leith.
\newblock {srsLTE: An Open-source Platform for {LTE} Evolution and
  Experimentation}.
\newblock In {\em Proc. of ACM WiNTECH}, pages 25--32, New York City, NY, USA,
  October 2016.

\bibitem{ivancic2019flying}
William~D Ivancic, Robert~J Kerczewski, Robert~W Murawski, Konstantin Matheou,
  and Alan~N Downey.
\newblock Flying drones beyond visual line of sight using 4g lte: Issues and
  concerns.
\newblock In {\em 2019 Integrated Communications, Navigation and Surveillance
  Conf. (ICNS)}, 2019.

\bibitem{izydorczyk2018experimental}
Tomasz Izydorczyk, M{\u{a}}d{\u{a}}lina Bucur, Fernando~ML Tavares, Gilberto
  Berardinelli, and Preben Mogensen.
\newblock Experimental evaluation of multi-antenna receivers for uav
  communication in live lte networks.
\newblock In {\em 2018 IEEE Globecom Workshops}, 2018.

\bibitem{khuwaja2018survey}
Aziz~Altaf Khuwaja, Yunfei Chen, Nan Zhao, Mohamed-Slim Alouini, and Paul
  Dobbins.
\newblock A survey of channel modeling for uav communications.
\newblock {\em IEEE Communications Surveys \& Tutorials}, 20(4):2804--2821,
  2018.

\bibitem{kitindi2017wireless}
Edvin~J Kitindi, Shu Fu, Yunjian Jia, Asif Kabir, and Ying Wang.
\newblock Wireless network virtualization with sdn and c-ran for 5g networks:
  Requirements, opportunities, and challenges.
\newblock {\em IEEE Access}, 5:19099--19115, 2017.

\bibitem{kumar2006medium}
Sunil Kumar, Vineet~S Raghavan, and Jing Deng.
\newblock Medium access control protocols for ad hoc wireless networks: A
  survey.
\newblock {\em Ad hoc networks}, 4(3):326--358, 2006.

\bibitem{larsen2018survey}
Line~MP Larsen, Aleksandra Checko, and Henrik~L Christiansen.
\newblock A survey of the functional splits proposed for 5g mobile crosshaul
  networks.
\newblock {\em IEEE Communications Surveys \& Tutorials}, 21(1):146--172, 2018.

\bibitem{lin2018sky}
Xingqin Lin, Vijaya Yajnanarayana, Siva~D Muruganathan, Shiwei Gao, Henrik
  Asplund, Helka-Liina Maattanen, Mattias Bergstrom, Sebastian Euler, and
  Y-P~Eric Wang.
\newblock The sky is not the limit: Lte for unmanned aerial vehicles.
\newblock {\em IEEE Magazine}, 56(4):204--210, 2018.

\bibitem{lyu2017blocking}
Jiangbin Lyu and Rui Zhang.
\newblock Blocking probability and spatial throughput characterization for
  cellular-enabled uav network with directional antenna.
\newblock {\em arXiv preprint arXiv:1710.10389}, 2017.

\bibitem{mei2019cellular}
Weidong Mei, Qingqing Wu, and Rui Zhang.
\newblock Cellular-connected uav: Uplink association, power control and
  interference coordination.
\newblock {\em IEEE Transactions on Wireless Communications}, 2019.

\bibitem{mei2019uplink}
Weidong Mei and Rui Zhang.
\newblock Uplink cooperative noma for cellular-connected uav.
\newblock {\em EEE J. Sel. Topics in Sig, Processing}, 13(3):644--656, 2019.

\bibitem{mogensen2007lte}
Preben Mogensen, Wei Na, Istv{\'a}n~Z Kov{\'a}cs, Frank Frederiksen, Akhilesh
  Pokhariyal, Klaus~I Pedersen, Troels Kolding, Klaus Hugl, and Markku Kuusela.
\newblock Lte capacity compared to the shannon bound.
\newblock In {\em 2007 ieee 65th vehicular technology Conf.-vtc2007-spring},
  pages 1234--1238. IEEE, 2007.

\bibitem{naqvi2018drone}
Syed Ahsan~Raza Naqvi, Syed~Ali Hassan, Haris Pervaiz, and Qiang Ni.
\newblock Drone-aided communication as a key enabler for 5g and resilient
  public safety networks.
\newblock {\em IEEE Communications Magazine}, 56(1):36--42, 2018.

\bibitem{nguyen2018ensure}
Huan~Cong Nguyen, Rafhael Amorim, Jeroen Wigard, Istv{\'a}n~Z Kov{\'a}cs,
  Troels~B S{\o}rensen, and Preben~E Mogensen.
\newblock How to ensure reliable connectivity for aerial vehicles over cellular
  networks.
\newblock {\em IEEE Access}, 6:12304--12317, 2018.

\bibitem{nikaein2014openairinterface}
Navid Nikaein, Raymond Knopp, Florian Kaltenberger, Lionel Gauthier, Christian
  Bonnet, Dominique Nussbaum, and Riadh Ghaddab.
\newblock {OpenAirInterface: an open LTE network in a PC}.
\newblock In {\em Proc. of Intl. Conf. on Mobile Computing and Networking},
  pages 305--308, Maui, HI, USA, September 2014.

\bibitem{polese2020experimental}
Michele Polese, Lorenzo Bertizzolo, Leonardo Bonati, Abhimanyu Gosain, and
  Tommaso Melodia.
\newblock An experimental mmwave channel model for uav-to-uav communications.
\newblock In {\em Proc. of 4th ACM Workshop on Millimeter-wave Networks and
  Sensing Systems (mmNets)}, London, UK, September 2020.

\bibitem{basinhopping}
Python.
\newblock scipy.optimize suite.

\bibitem{LP0965}
Ettus Reseach.
\newblock {LP0965}, 2020.
\newblock {https://kb.ettus.com/images/0/06/
  ettus\_research\_lp0965\_datasheet.pdf}.

\bibitem{vert2450}
Ettus Reseach.
\newblock {Vert2450}, 2020.
\newblock {https://kb.ettus.com/images/9/9e
  /ettus\_research\_vert2450\_datasheet.pdf}.

\bibitem{shell}
Shell.
\newblock Eye in the sky, 2019.

\bibitem{sheshadri2020skyhaul}
Ramanujan~K Sheshadri, Eugene Chai, Karthikeyan Sundaresan, and Sampath
  Rangarajan.
\newblock Skyhaul: An autonomous gigabit network fabric in the sky.
\newblock {\em arXiv preprint arXiv:2006.11307}, 2020.

\bibitem{stanczak2018mobility}
Jedrzej Stanczak, Istvan~Z Kovacs, Dawid Koziol, Jeroen Wigard, Rafhael Amorim,
  and Huan Nguyen.
\newblock Mobility challenges for unmanned aerial vehicles connected to
  cellular lte networks.
\newblock In {\em 2018 IEEE 87th Vehicular Technology Conf. (VTC Spring)},
  2018.

\bibitem{sundaresan2018skylite}
Karthikeyan Sundaresan, Eugene Chai, Ayon Chakraborty, and Sampath Rangarajan.
\newblock Skylite: End-to-end design of low-altitude uav networks for providing
  lte connectivity.
\newblock {\em arXiv preprint arXiv:1802.06042}, 2018.

\bibitem{van2016lte}
Bertold Van~der Bergh, Alessandro Chiumento, and Sofie Pollin.
\newblock Lte in the sky: Trading off propagation benefits with interference
  costs for aerial nodes.
\newblock {\em IEEE Communications Magazine}, 54(5):44--50, 2016.

\bibitem{vilzmann2006survey}
Robert Vilzmann and Christian Bettstetter.
\newblock A survey on mac protocols for ad hoc networks with directional
  antennas.
\newblock In {\em EUNICE 2005: Networks and Applications Towards a Ubiquitously
  Connected World}, pages 187--200. Springer, 2006.

\bibitem{zeng2018cellular}
Yong Zeng, Jiangbin Lyu, and Rui Zhang.
\newblock Cellular-connected uav: Potential, challenges, and promising
  technologies.
\newblock {\em IEEE Wireless Communications}, 26(1):120--127, July 2018.

\end{thebibliography}

\bibliographystyle{plain}
\vspace{-1cm}
\begin{IEEEbiography}[{\includegraphics[width=1in,height=1.25in,clip,keepaspectratio]{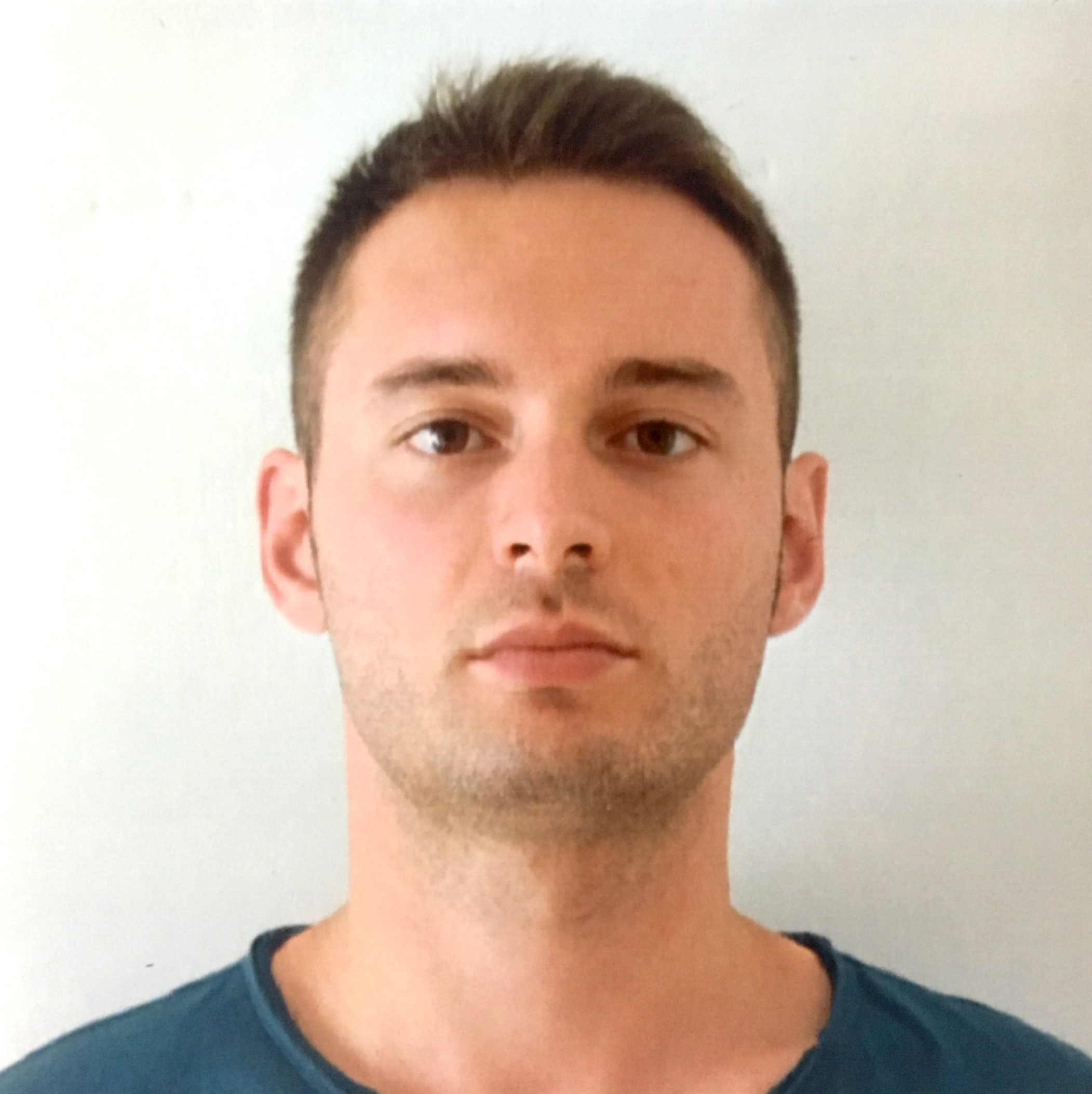}}]{Lorenzo Bertizzolo} 
earned his Ph.D. in Computer Engineering at the Institute for the Wireless Internet of Things at Northeastern University, under Prof. Tommaso Melodia in 2021. Previously, he earned his B.S. and his M.S. in Computer and Communication Networks Engineering with honors from Politecnico di Torino, Italy in 2014 and 2015, respectively.
His research focuses on 5G, software-defined networking for wireless, network optimization, and non-terrestrial UAV networks. 
He is a collaborator of AT\&T Labs Research, working on the integration of Unmanned Aerial System into the next generations' cellular networks. 
He received the Best Paper Award at the ACM Workshop on Wireless Network Testbeds, Experimental evaluation \& CHaracterization (WiNTECH) in 2019 and the ECE Outstanding Research Award from Northeastern University in 2020. In 2021, he joined Apple as a Wireless Systems Engineer.
\end{IEEEbiography}
\vspace{-1cm}
\begin{IEEEbiography}[{\includegraphics[width=1in,height=1.25in,clip,keepaspectratio]{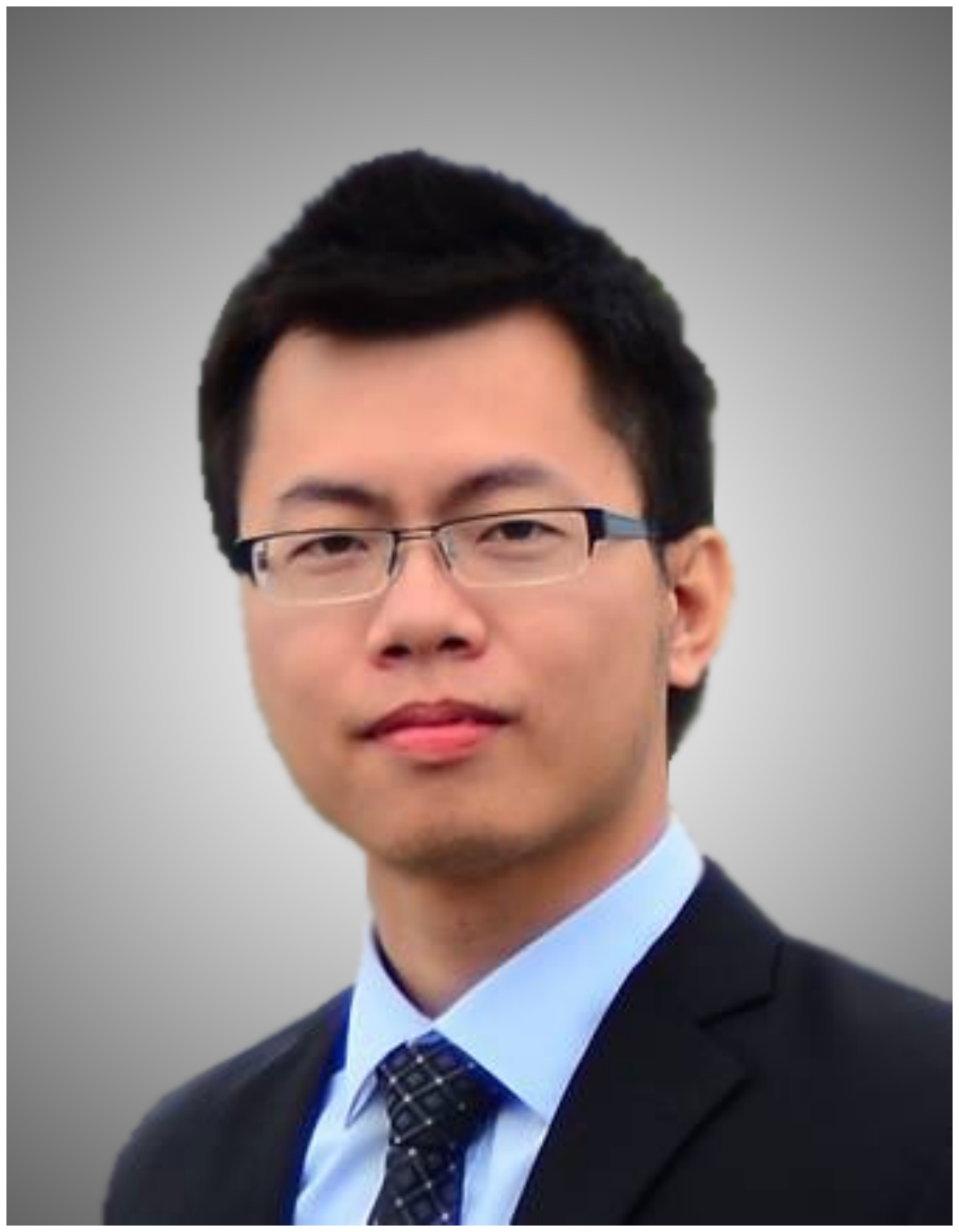}}]{Tuyen X. Tran} received the BEng degree (honors program) in electronics and telecommunications from the Hanoi University of Science and Technology, Vietnam, in 2011, the MSc degree in ECE from the University of Akron, Ohio, in 2013, and the PhD degree in electrical and computer engineering (ECE) from Rutgers University, New Jersey, in 2018. His research interests include wireless communications, mobile cloud computing, and network optimization. He is currently a senior inventive scientist with AT\&T Labs Research, Bedminster, New Jersey. He received the Best Paper Award at the IEEE/IFIP Wireless On-demand Network Systems and Services Conference (WONS) in 2017 and the Outstanding Graduate Student Award from the Rutgers School of Engineering in 2018. He is a member of the IEEE and ACM.
\end{IEEEbiography}
\vspace{-1cm}

\begin{IEEEbiography}[{\includegraphics[width=1in,height=1.25in,clip,keepaspectratio]{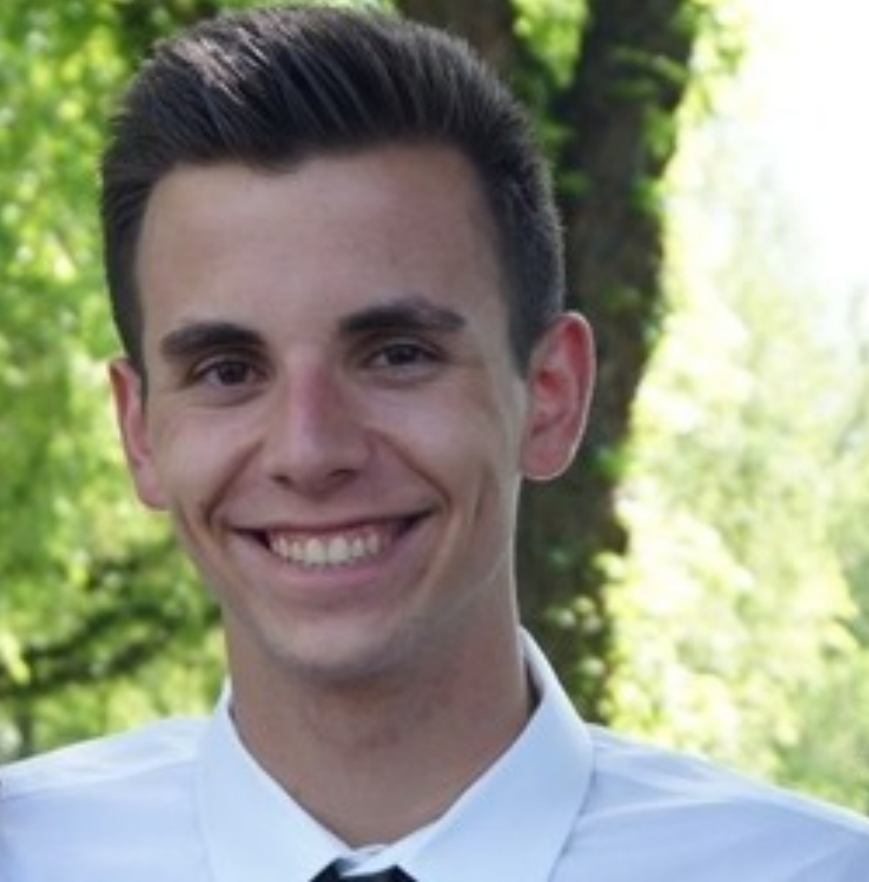}}]{John Buczek} is currently pursuing a BSMS in Electrical Engineering with concentration in Power Systems at Northeastern University, Boston, MA. He is an undergraduate research assistant at the Institute for the Wireless IoT at Northeastern University. His research interests include Power Electronics, Unmanned Aerial Vehicles and UAV Networks, and the Internet of Things.
\end{IEEEbiography}
\vspace{-1cm}
\begin{IEEEbiography}[{\includegraphics[width=1in,height=1.25in,clip,keepaspectratio]{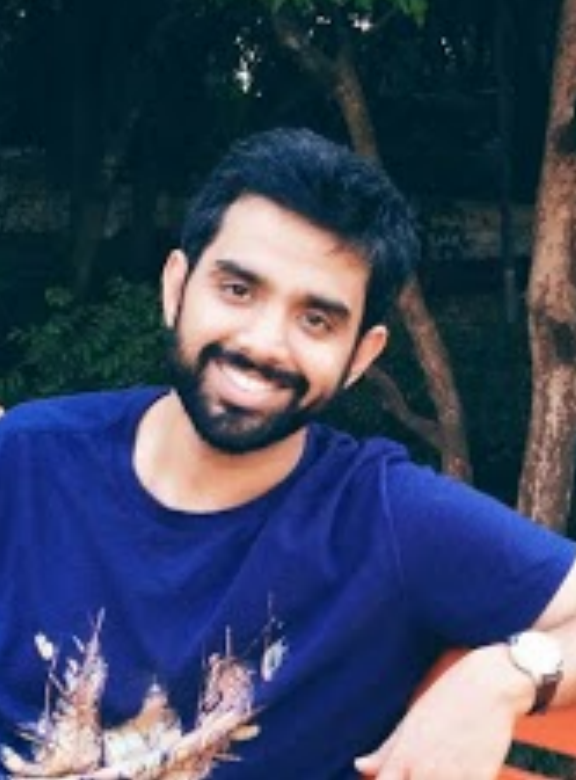}}]{Bharath Balasubramanian} is a Principal Inventive Scientist in the Network Infrastructure Research department at AT\&T Labs Research, Bedminster, New Jersey. Prior to this, he received his Ph.D. from the University of Texas at Austin and Bachelors in Electronics Engineering (B.E) from Mumbai University. His research interests include distributed coordination, consistency fault tolerance and state management in large scale distributed systems.
\end{IEEEbiography}
\vspace{-1cm}
\begin{IEEEbiography}[{\includegraphics[width=1in,height=1.25in,clip,keepaspectratio]{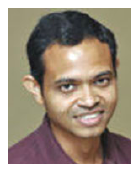}}]{Rittwik Jana} is currently the Chief architect of RAN Intelligence at VMware. Prior to this, he worked at AT\&T Labs Research for 21 years. His interests span architecting the disaggregated RAN intelligent controller in O-RAN, service composition of VNFs using TOSCA, model driven control loop and automation in ONAP, video streaming for cellular networks and systems. He is a recipient of the AT\&T Science and Technology medal in 2016 for contributions to model driven cellular network planning, the IEEE Jack Neubauer memorial award in 2017 for systems work on full duplex wireless and several best paper awards in wireless communications. He holds 100 granted patents and has published over 150 technical papers at IEEE and ACM conferences and journals.

\end{IEEEbiography}
\vspace{-1cm}
\begin{IEEEbiography}[{\includegraphics[width=1in,height=1.25in,clip,keepaspectratio]{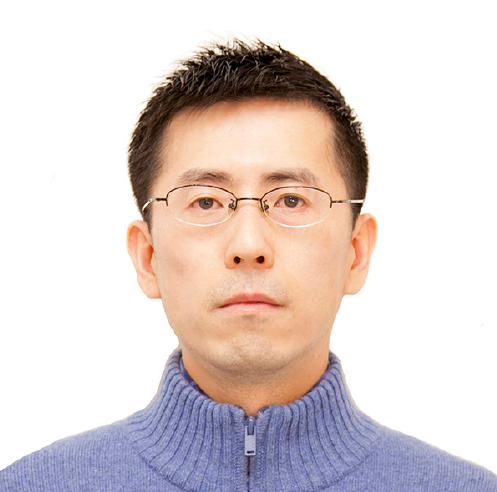}}]{Yu Zhou} received his Ph.D. in Computer Science from The George Washington University and B.S. in Computer Science from Peking University, China. Currently, he is a Principal Inventive Scientist at AT\&T Labs Research. He is an inventor of about 90 patents in the wireless communications. His research interests include graph theory and network optimization algorithms.
\end{IEEEbiography}
\vspace{-1cm}

\begin{IEEEbiography}[{\includegraphics[width=1in,height=1.25in,clip,keepaspectratio]{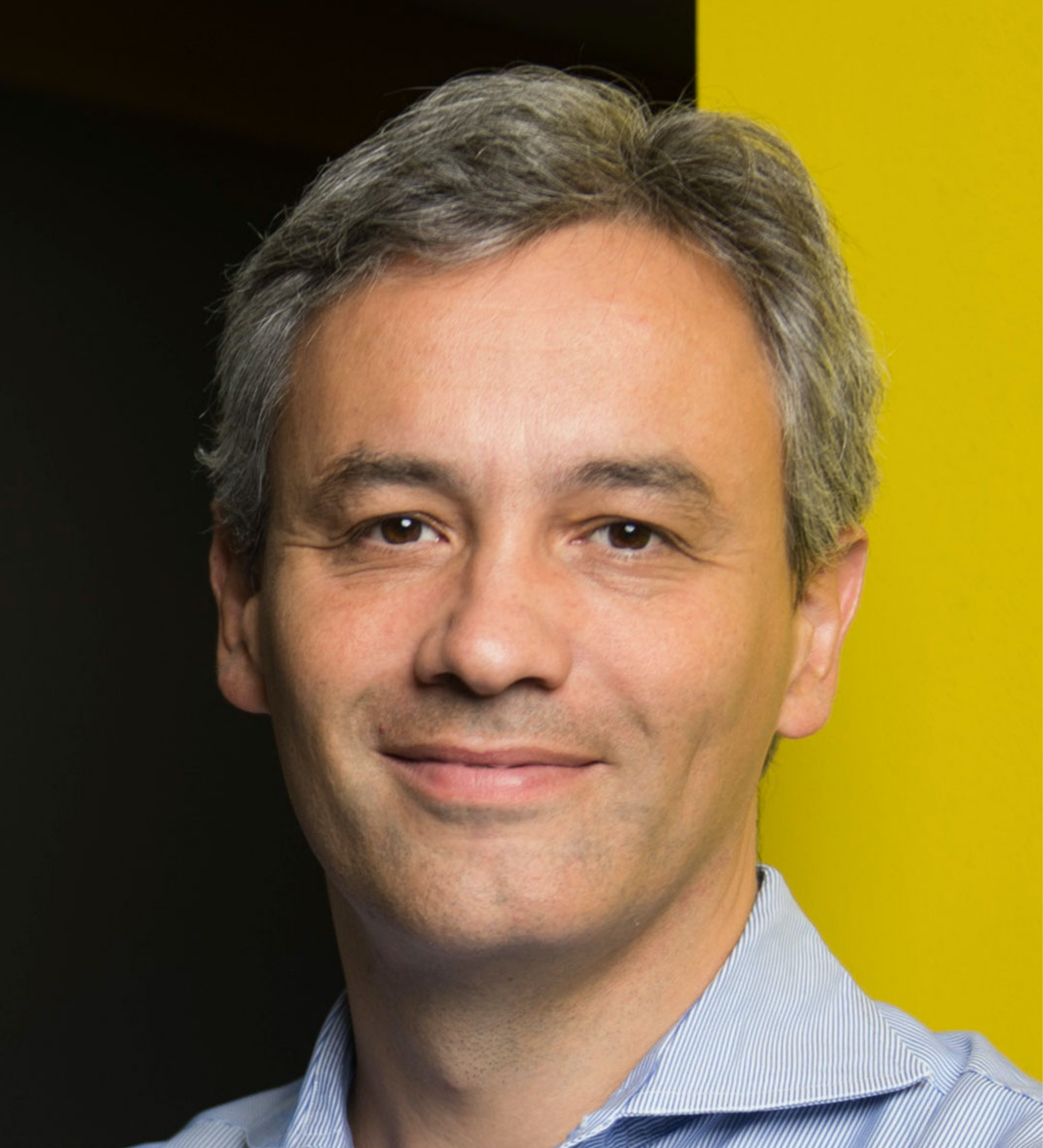}}]{Tommaso Melodia} [F’18] received the Ph.D. degree in electrical and computer engineering from the Georgia Institute of Technology in 2007.
He is currently the William Lincoln Smith Professor
with the Department of Electrical and Computer
Engineering, Northeastern University. He is also the
Director of the Institute for the Wireless Internet of
Things, and the Director of Research for the PAWR
Project Office, a public-private partnership that is
developing four city-scale platforms for advanced
wireless research in the United States. His research
focuses on modeling, optimization, and experimental evaluation of wireless networked systems, with applications to 5G networks and Internet of Things, software-defined networking, and body area networks. His research is supported mostly by the U.S. federal agencies, including the National
Science Foundation, the Air Force Research Laboratory, the Office of Naval Research, the Army Research Laboratory, and DARPA. He is a Senior Member of the ACM. He is the Editor-in-Chief of Computer Networks, and a former Associate Editor of the IEEE TRANSACTIONS ON WIRELESS COMMUNICATIONS, the IEEE TRANSACTIONS ON MOBILE COMPUTING, the IEEE TRANSACTIONS ON MULTIMEDIA, among others.
\end{IEEEbiography}




\end{document}